\documentstyle[12pt,a41,eucal,epsfig,axodraw,cite,amsmath]{article}
 \newcommand{\PO}{\hat{P}}
 
 \newcommand{\N}{\nonumber}
 \newcommand{\ep}{\varepsilon}

\newcommand{\bea}{\begin{eqnarray}}
\newcommand{\bq}{\begin{equation}}
\newcommand{\eea}{\end{eqnarray}}
\newcommand{\eq}{\end{equation}}

\newcommand{\gsim}{\raisebox{-0.07cm   }
{$\, \stackrel{>}{{\scriptstyle\sim}}\, $}}

\newcommand\epsi{\varepsilon}

\newcommand\be{\begin{eqnarray}}
\newcommand\ee{\end{eqnarray}}

\begin{document}
\noindent
\sloppy
\thispagestyle{empty}
\begin{flushleft}
DESY 08-029 \hfill
\\
SFB/CPP-08-16\\
SFB F013 Technical Report no. 2008-04 \\
February 2008
\end{flushleft}
%
\vspace*{\fill}
\hspace{-3mm}
{\begin{center}
{\LARGE\bf Two--Loop Massive Operator Matrix Elements}

\vspace{2mm}
{\LARGE\bf 
\boldmath 
for~Unpolarized~Heavy~Flavor~Production~to~$O\!(\epsilon)$}
\end{center}
}

\begin{center}
\vspace{2cm}
\large
Isabella Bierenbaum, Johannes Bl\"umlein, Sebastian Klein
\\
\vspace{5mm}
\normalsize
{\it Deutsches Elektronen--Synchrotron, DESY,\\
Platanenallee 6, D--15738 Zeuthen, Germany}
\\

\vspace{7mm}
\large
Carsten Schneider
\\
\vspace{5mm}
\normalsize
{\it Research Institute for Symbolic Computation (RISC),\\
Johannes Kepler University, Altenbergerstra\ss{}e 69, A-4040 Linz, Austria}
\\
\vspace{2em}
\end{center}
\vspace*{\fill}
%
\begin{abstract}
\noindent
We calculate the $O(\alpha_s^2)$ massive operator matrix elements for the 
twist--2 operators, which contribute to the heavy flavor Wilson coefficients
in unpolarized deeply inelastic scattering in the region $Q^2 \gg m^2$, up to 
the $O(\varepsilon)$ contributions. These terms contribute through the 
renormalization of the $O(\alpha_s^3)$ heavy flavor Wilson coefficients of the 
structure function $F_2(x,Q^2)$. The calculation has been performed using 
light--cone expansion techniques without using the integration-by-parts method. 
We represent the individual Feynman diagrams by generalized hypergeometric 
structures, the $\varepsilon$--expansion of which leads to infinite sums depending
on the Mellin variable $N$. These sums are finally expressed in terms of nested 
harmonic sums using the general summation techniques implemented in the 
{\tt Sigma} package. 
\end{abstract}
\vspace*{\fill}
\newpage
\section{Introduction}
\label{sec:introduction}

\vspace{1mm}\noindent
The heavy flavor corrections to deeply inelastic scattering constitute an
important part of the structure functions in the lower $x$ region, 
cf. \cite{THOMP}. 
The current world data for the nucleon structure functions $F_2^{p,d}(x,Q^2)$
reached the precision of a few per cent over a wide kinematic region. 
Therefore both for the determination of the QCD scale $\Lambda_{\rm QCD}$
and the detailed shapes of the partonic distribution functions the analysis
at the level of the $O(\alpha_s^3)$ corrections is required to control the
theory-errors on the level of the experimental accuracy and below \cite{ALP}. 
In a recent non--singlet analysis \cite{BBG} errors for $\alpha_s(M_Z^2)$ of $O(1.5~\%)$
were obtained extending the analysis effectively to N$^3$LO. In the flavor singlet case
the yet unknown 3--loop heavy flavor Wilson coefficients prevent a consistent 3--loop analysis.
Due to the large statistics in the lower $x$ region one may hope to eventually improve the
accuracy of $\alpha_s(M_Z^2)$ beyond the above value. 

The heavy flavor corrections to $F_2^{p,d}(x,Q^2)$ were calculated to 2--loop order
in the whole kinematic domain in a semi-analytic way in $x$-space in Refs.~\cite{HEAV1}. 
A fast implementation for complex $N$--space was given in \cite{HEAV2}. In the 
range 
of higher values of $Q^2$ one may calculate the heavy flavor Wilson coefficients
to the structure functions $F_2(x,Q^2)$ and $F_L(x,Q^2)$ in analytic form.
For $F_2(x,Q^2)$ this calculation has been performed to 2--loop order in 
\cite{BUZA,BBK1} and for $F_L(x,Q^2)$ to 3--loop order in \cite{BFNK}. 
In the region $Q^2 \gg m^2$ the heavy flavor Wilson coefficients for deep--inelastic 
scattering factorize into massive operator matrix elements $A_{ij}(\mu^2/m^2)$ and 
the massless Wilson coefficients $C_k(Q^2/\mu^2)$ \cite{WIL1,WIL2,WIL3} for all but 
the power suppressed contributions. The massive operator matrix elements are 
universal and contain all the mass dependence in the logarithmic orders and the 
constant term. The process dependence is due to the massless Wilson 
coefficients.
In the case of the structure function $F_2(x,Q^2)$ the asymptotic heavy flavor 
contributions become quantitatively very close to those obtained in the complete 
calculation \cite{CLO,HEAV1} at LO and NLO already for $Q^2 \gsim 10~m^2$. 
These
scales are sufficiently low and match with the region analyzed in deeply inelastic 
scattering. 

In the present paper we perform a first step towards the 3--loop heavy flavor Wilson 
coefficients for the structure function $F_2(x,Q^2)$. The renormalization of the 
massive operator matrix elements to 3--loop order encounters also the contributions
of $O(\ep)$ at $O(\alpha_s^2)$, which have not yet been calculated before.~\footnote{In 
the massless case the off--shell operator matrix elements were calculated to this
order for space--like momenta in the $\overline{\rm MS}$--scheme 
for unpolarized and polarized deeply inelastic scattering in \cite{OM1,OM2}, 
which are needed in the 
calculation of the 3--loop anomalous dimensions.} The 2--loop $O(\ep)$ terms
form finite contributions to the $O(a_s^3)$ matrix elements with the 
single pole terms emerging at 1st order.
We extend the work
presented previously in Ref.~\cite{BBK1}. For the calculation of the $O(\ep)$ 
2--loop contributions our representation which is based on hypergeometric integrals
was extended straightforwardly. However, many more infinite nested sums, which contain 
the Mellin variable $N$, had to be evaluated for the first time, since other 
available techniques \cite{SUMMER,NESTS,XSUMMER} could not be used for this 
purpose. We applied both
suitable integral representations and the summation package {\tt Sigma} 
\cite{sigma},
which solves these sums in $\Pi\Sigma$--fields. In the result all sums can be 
expressed in terms of nested harmonic sums \cite{BK1,SUMMER}.

The paper is organized as follows. In section~2 the structure of the heavy 
flavor contributions to the deeply inelastic structure function is summarized
for the kinematic region $Q^2 \gg m^2$. The renormalization of the massive 
operator matrix elements to 3--loop order is described in section~3. In 
section~4 the $O(\ep)$ contributions to the 2--loop operator matrix elements
are calculated. Section~5 contains the conclusions. In the appendices we 
present details of the calculation, newly derived infinite sums and related 
functions depending on the Mellin parameter $N$, and
a further check on our result comparing the Abelian part of the first 
moment with the corresponding part of the on--shell photon propagator, 
\section{Basic Formalism}
\label{sec:basform}

\vspace{1mm}\noindent
In the twist--2 approximation, the deep--inelastic nucleon structure functions 
$F_n(x,Q^2),~n=2,L,$
are described as Mellin convolutions between the parton densities 
$f_j(x,\mu^2)$ and the Wilson coefficients ${\mathsf{C}}_i^j(x,Q^2/\mu^2)$
\begin{eqnarray}
\label{STR}
F_n(x,Q^2) &=& \sum_j {\mathsf{C}}_n^j\left(x,\frac{Q^2}{\mu^2}\right) \otimes 
f_j(x,\mu^2)
\end{eqnarray}
to all orders in perturbation theory due to the factorization theorem. 
Here $\mu^2$ denotes the factorization scale and
the Mellin convolution is given by the integral
\begin{eqnarray}
[A \otimes B](x) = \int_0^1 dx_1 \int_0^1 dx_2~~ \delta(x - x_1 x_2) 
~A(x_1) B(x_2)~.
\end{eqnarray}
The distributions $f_j$ refer to {\sf massless} partons and the heavy 
flavor effects are contained in the Wilson coefficients only.
As was shown in Ref.~\cite{BUZA} in the region $Q^2 \gg 
m^2$ all non--power contributions to the heavy quark Wilson coefficients
obey
\begin{eqnarray}
\label{HFAC}
H_{n;i}^{\sf fl}\left(\frac{Q^2}{m^2}, \frac{m^2}{\mu^2},x\right) = 
C_{n;k}^{\sf fl} \left(\frac{Q^2}{\mu^2},x\right) 
\otimes	A_{k,i}^{\sf fl} \left(\frac{m^2}{\mu^2},x\right)~, 
\end{eqnarray}
where $C_{n;k}^{\sf fl} \left({Q^2}/{\mu^2},x\right)$ are the Wilson 
coefficients for massless partons and  
$A_{k,i}^{\sf fl} \left({m^2}/{\mu^2},x\right)$ are the massive 
operator matrix elements. Here $\mu$ refers to the factorization scale between 
the heavy and light contributions in ${\mathsf{C}}^j_n$.
In the convolution (\ref{HFAC}) only those terms are 
accounted for which contribute to the respective heavy flavor Wilson coefficient 
functions. The index {\sf fl} denotes the 
flavor-decomposition and labels the pure--singlet and gluon contributions 
(PS,G) and three non-singlet (NS$^{\pm}$, NS$^{\rm v}$) combinations. 
Due to the fact that the diagrams considered here contain one heavy quark 
line, at $O(a_s)$ only $A_{Qg}$ contributes. Beginning with $O(a_s^2)$   
there is also the pure-singlet  $A_{Qq}^{\rm PS}$ and 
the non-singlet term $A_{Qq}^{\rm NS^+}$, while at $O(a_s^3)$ also the two 
other non-singlet terms $A_{Qq}^{\rm NS^-}$ and $A_{Qq}^{\rm NS^v}$ 
contribute. The corresponding combinations of quark distributions  
for the singlet and non-singlet terms are $\Sigma(x,Q^2)$ and
$q^{\rm NS^l}(x,Q^2)$ with
\begin{eqnarray}
\Sigma(x,Q^2) &=& \sum_{k=1}^{N_l} \left[q_k(x,Q^2) - 
\overline{q}_k(x,Q^2)\right]\\
q^{\rm NS^{\pm}}_{mn}(x,Q^2) &=& 
\left[q_m(x,Q^2) \pm \overline{q}_m(x,Q^2)\right]
-\left[q_n(x,Q^2) \pm \overline{q}_n(x,Q^2)\right]\\
q^{\rm NS^v}(x,Q^2) &=& \sum_{k=1}^{N_f} \left[q_k(x,Q^2) - 
\overline{q}_k(x,Q^2)\right]~.
\end{eqnarray}
$N_l$ denotes the number of light quark flavors. The massless Wilson 
coefficients were calculated in
\cite{WIL1,WIL2,WIL3} to 3--loop orders. The massive operator matrix elements
are process independent quantities.
The factorization (\ref{HFAC}) is a 
consequence of the renormalization group equation.
In Mellin space the operator matrix elements $A_{k,i}^{\sf fl}$ and 
light flavor Wilson coefficients 
obey the following expansions~:
\begin{eqnarray}
\label{op1}
A_{k,i}^{\rm fl} \left(\frac{m^2}{\mu^2}\right) = \langle 
i|O_k|i\rangle
= \delta_{k,i} + \sum_{l=1}^{\infty} a_s^l A_{k,i}^{{\rm fl},(l)},~~~~i=q,g 
\\
\label{coeQ}
C_{2;i}^{\rm fl} \left(\frac{Q^2}{\mu^2}\right)
= \delta_{i,q} + \sum_{l=1}^{\infty} a_s^l C_{2,i}^{{\rm fl},(l)},~~~~i=q,g 
\label{coeg}
\end{eqnarray}
of the twist--2 flavor singlet,  non--singlet and gluon operators $O_k^{\rm 
NS,S,g}$ between {\sf partonic} states $|i\rangle$, 
which are related by collinear factorization to the initial--state nucleon 
states $|N\rangle$. The local operators are given by
\begin{eqnarray}
\label{op2}
O_{q,r}^{{\rm NS}, \mu_1, \ldots, \mu_N}(z) &=& \frac{1}{2} i^{N-1}  
S \left[\overline{q}(z) \gamma^{\mu_1}
D^{\mu_2} \ldots D^{\mu_N} \frac{\lambda_r}{2}
q(z)\right] - {\sf Trace~Terms}
\\
O_q^{{\rm S}, \mu_1, \ldots, \mu_N}(z) &=& \frac{1}{2} i^{N-1}  
S \left[\overline{q}(z) \gamma^{\mu_1} D^{\mu_2} 
\ldots D^{\mu_N} q(z)\right] - {\sf Trace~Terms}
\\
O_g^{\mu_1, \ldots, \mu_N}(z) &=& \frac{1}{2} i^{N-2}  
S \left[F_{\alpha}^{a,\mu_1}(z) D^{\mu_2} 
\ldots D^{\mu_{N-1}} F^{a,\alpha, \mu_N}(z)\right] - {\sf Trace~Terms}~.
\end{eqnarray}
Here $S$ denotes the operator which symmetrizes all Lorentz-indices and
$D_{\mu_1} = \partial_{\mu_1}  - g  t_a A^a_{\mu_1}$ is the covariant
derivative, $q(z)$, $\overline{q}(z)$ and $F^{a,\mu\nu}(z)$ denote the
quark-, anti-quark field and the gluon field-strength operators,
with $g = (4 \pi\alpha_s)^{1/2} = (16 \pi^2 a_s)^{1/2}$ 
the strong coupling constant, $t_a$ the generators of
$SU(3)_c$, and $\lambda_r$ the Gell-Mann matrices of $SU(3)_F$.
The Feynman rules for the operator insertions are given in \cite{BBK1,YND}. 
\section{\bf\boldmath Renormalization of the Matrix Elements}
\label{REN}

\vspace{1mm}\noindent
The massive operator matrix elements contain ultraviolet and collinear 
singularities which have to be renormalized. Charge-, mass-,  operator-, and wave 
function renormalization have to be performed. 
Collinear singularities appear in those parts of the  diagrams with vertices 
which link only to massless lines, and are specific to the particular classes of 
diagrams.  Since in the present case at least one closed fermion line is massive, 
collinear singularities appear only at $O(a_s^2)$. 
The un-renormalized massive operator matrix elements read
\begin{eqnarray}
\hat{\hat{A}}_{ij} = \delta_{ij} + \sum_{k=0}^\infty \hat{a}_s^k \hat{\hat{A}}_{ij}^{(k)}~.
\end{eqnarray}
Here $\hat{a}_s$ denotes the bare coupling constant.
To 2--loop order, the corresponding diagrams were given in Ref.~\cite{BUZA}. Here one has to 
distinguish one-particle irreducible and reducible diagrams, which both contribute in the 
calculation. We would like to remind the reader the background of this aspect. 

If one evaluates the heavy-quark Wilson coefficients in an usual Feynman-diagram 
calculation, the matrix elements are given by diagrams of the type depicted in Figure~1. 
\begin{figure}[hbt]
\begin{center}
\includegraphics[angle=0, width=5.0cm]{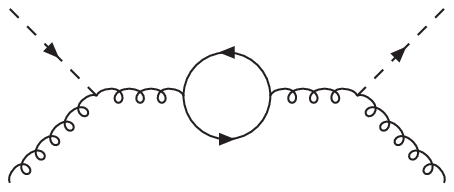} 
\end{center}
\caption{\label{fig:1}
\sf Massive quark self-energy correction to virtual scalar--gluon scattering}
\end{figure}
The incoming gluon is factorized from the nucleon, i.e. we assume the parton 
life-time $\tau_L$ being much longer than the interaction time $\tau_I$ of the 
virtual photon 
with the nucleon. As is well-known \cite{DY}, this condition is fulfilled whenever 
$k_{\perp}^2 \ll Q^2$ and neither the Bjorken variable $x$ is very small ($x \ll 
\hspace{-4mm} / \hspace{3mm} 1$) nor large ($x \approx \hspace{-4mm} / \hspace{3mm} 1$). 
This is the case performing the Bjorken limit and applying the collinear parton model, 
in which the incoming massless partons are dealt with as on--shell particles. 
Also in this case, self-energy diagrams for the incoming parton lines are present. 
However, one may 
factorize these contributions into the {\sf non--perturbative} parton densities at leading 
twist, resp. parton correlation functions at higher twist, since these contributions are 
{\sf virtual} and are always present whatever hard scattering cross section is considered. 
They do not form a heavy quark signature which can be identified in a 
subspace of the complete 
final--state Fock--space emerging in deeply inelastic lepton--nucleon scattering. This 
procedure was adopted in Ref.~\cite{HEAV1}. One consequence is that at $O(a_s^2)$ there
are no diagrams with two fermion lines, resp. at $O(a_s^3)$ none with three fermion 
lines in the
general heavy flavor Wilson coefficients. The 
situation is different in case of the operator matrix elements obtained after the 
light--cone expansion is being performed. Here, the line between the two virtual photon- or weak 
gauge boson vertices is contracted. This line may contain virtual corrections, 
see e.g. Figure~1, which would be 
lost in the process of contraction. They have to be accounted for in attaching these 
self--energies to outer lines of the contracted diagram, see Figure~2. 
From the case of the fermion--fermion anomalous dimension at leading order 
these aspects are known for long \cite{FERMI,GW,LOSP}.  
\begin{figure}[hbt]
\begin{center}
\includegraphics[angle=0, width=9.0cm]{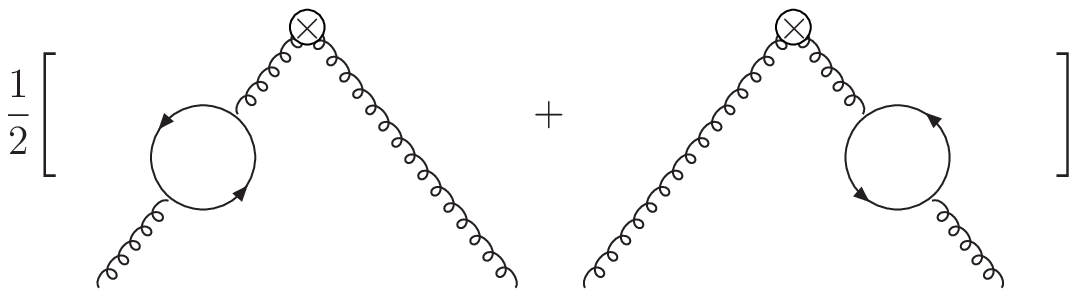} 
\end{center}
\caption{\label{fig:2}
\sf $O(a_s)$ Self-energy correction due to massive quarks for the operator matrix element
$A_{gg}^{(1)}$. 
}
\end{figure}

In kinematic regions, where higher twist effects can be safely neglected~\cite{BBG,BB} and 
at sufficiently high scales $Q^2$, the scaling violations of deeply inelastic 
structure functions are due to the running coupling constant and heavy quark
mass effects, after target mass effects \cite{GP} have been accounted for. We will 
further assume that we are in a region where power corrections due to heavy
quarks are negligibly small, i.e., the heavy quark effects contribute 
logarithmically $\propto \ln^l(m^2/\mu^2),~~l \geq 0$. In this region one
may express the structure functions $F_i(N,Q^2)$ in Mellin space by 
\begin{eqnarray}
F_i(N,Q^2) &=& \sum_{l=1}^{N_l} C_{i,q}(N,Q^2/\mu_f^2;a_s(\mu_r^2)) \cdot 
\left[q_l(N,\mu_f^2/\mu_0^2,a_s(\mu_r^2)) + 
\overline{q}_l(N,\mu_f^2/\mu_0^2,a_s(\mu_r^2))\right] \nonumber \\ & &
+ C_{i,g}(N,Q^2/\mu_f^2;a_s(\mu_r^2)) \cdot g(N,\mu_f^2/\mu_0^2,a_s(\mu_r^2)) 
\nonumber
\end{eqnarray} \begin{eqnarray}
& & 
+ \sum_{h=1}^{N_l} H_{i,q}(N,Q^2/\mu_f^2;a_s(\mu_r^2)) \cdot 
\left[q_l(N,\mu_f^2/\mu_0^2,a_s(\mu_r^2)) + 
\overline{q}_l(N,\mu_f^2/\mu_0^2,a_s(\mu_r^2))\right] \nonumber\\ & &
+ H_{i,g}(N,Q^2/\mu_f^2;a_s(\mu_r^2)) \cdot g(N,\mu_f^2/\mu_0^2,a_s(\mu_r^2))~. 
\end{eqnarray}
Here $\mu_r$ and $\mu_f$ denote the renormalization and factorization scales, 
respectively, $\mu_0$ is a hadronic scale, and $q, \overline{q}$ and $g$ 
denote the quark- and gluon distribution functions.  Furthermore, the heavy 
quark Wilson coefficients factorize according to (\ref{HFAC}), which is 
described by the scale $\mu$. In the following we identify all these scales
$\mu = \mu_r = \mu_f$. Since the structure functions $F_i(N,Q^2)$ do not 
depend on these scales they obey the following
renormalization group equation (RGE) \cite{RGE} 
\begin{eqnarray}
\label{ren1}
\left[{\cal D} + 2 \gamma_j\right]~F_n(N,Q^2) =0~,
\end{eqnarray}
where the differential operator ${\cal D}$ is defined by
\begin{eqnarray}
{\cal D} =
\mu^2 \frac{\partial}{\partial \mu^2} 
+ \beta(a_s(\mu^2)) \frac{\partial}{\partial 
a_s(\mu^2)} 
- \gamma_m(a_s(\mu^2)) m^2(\mu^2)\frac{\partial}{\partial m^2(\mu^2)}~.
\end{eqnarray}
$\beta(a_s)$ denotes the $\beta$-function, $\gamma_m(a_s)$ the mass 
anomalous dimension, and $\gamma_j$ denote the anomalous dimensions of the quark fields. 
Here we discuss the case of conserved currents, which have vanishing 
anomalous dimensions.~\footnote{Calculating the evolution of the transversity 
structure function $h_1(x,Q^2)$ using the forward Compton amplitude, this is 
not the case, cf.~\cite{H1}.}

The RGE for the Wilson coefficients and the parton distributions read 
\cite{RGE1,GW}
\begin{eqnarray}
\left[{\cal D}~\delta_{kj} - \gamma^N_{kj}(a_s)\right] 
{\mathsf{C}}_n^j(N,Q^2/\mu^2) &=& 0 \\
\left[\left({\cal D} + 2 \gamma_j\right) \delta_{kk'} + 
\gamma^N_{kk'}(a_s)\right] 
{{f}}_{jk'}(N,\mu^2/\mu_0^2) &=& 0~, 
\end{eqnarray}
with
\begin{eqnarray}
f_{jk}(N,\mu^2/\mu_0^2) = \langle j| O_k |j \rangle~. 
\end{eqnarray}
In the following we describe the renormalization to $O(a_s^3)$.
\subsection{Charge Renormalization}

\vspace{1mm} \noindent
We perform the charge renormalization in the $\overline{\rm MS}$-scheme. This 
allows to compare the results obtained in the QCD analysis of deeply inelastic 
scattering data with analyzes of other data. The bare coupling constant 
$\hat{a}_s$ is expressed by 
the renormalized coupling $a_s$  in the $\overline{\rm MS}$ scheme by 
\begin{eqnarray}
\hat{a}_s(\ep)        &=& Z_g^2(\ep,\mu^2) a_s(\mu^2) \nonumber\\
                 &=& a_s(\mu^2)\left[1 
+ \delta a_{s, 1} a_s(\mu^2)                
+ \delta a_{s, 2} a_s^2(\mu^2)   \right] + O(a_s^4)\\
\delta a_{s, 1} &=& S_\ep \frac{2 \beta_0}{\ep} \\
\delta a_{s, 2} &=& S_\ep^2 \left[\frac{4 \beta_0^2}{\ep^2} + 
\frac{\beta_1}{\ep}\right]~,
\end{eqnarray}
with $Z_g$ the $Z$--factor for the strong charge. 
Here the spherical factor $S_\ep$ is given by
\begin{eqnarray}
S_\ep = \exp \left[ \frac{\ep}{2}\left(\gamma_E - \ln(4\pi)\right)\right]~,
\end{eqnarray}
with $\gamma_E$ the Euler--Mascheroni number, $\ep = D - 4$, and $D$ the 
dimension of space--time. $\beta_0$ and $\beta_1$ 
\cite{B01} denote the first expansion coefficients of the $\beta$-function in 
the massless case
\begin{eqnarray}
\frac{d a_s(\mu^2)}{d \ln(\mu^2)} 
&=& \frac{1}{2} \ep a_s(\mu^2) - \sum_{k=0}^\infty \beta_k a_s^{k+2}(\mu^2)
\\
\beta_0 &=& \frac{11}{3} C_A - \frac{4}{3} T_F n_f \\
\beta_1 &=& \frac{34}{3} C_A^2 - 4 \left(\frac{5}{3} C_A + C_F\right) T_F n_f~. 
\end{eqnarray}
The color factors for $SU(3)_c$
are $C_F = (N_c^2-1)/(2 N_c) = 4/3, C_A = N_c = 3, T_F = 1/2$.
$n_f$ denotes the number of active flavors. The renormalized coupling constant
is obtained absorbing $Z_g$ into the bare coupling $\hat{g}$. 

In \cite{BUZA} a slightly 
different point of view was taken, including mass effects in the evolution of
$a_s(\mu^2)$ which usually means to choose another scheme, as e.g. the 
MOM-scheme \cite{MOM}. To maintain the Slavnov-Taylor identities of QCD the
calculation has to be performed using the background-field method \cite{BGF} 
in \cite{RUNA}.~\footnote{Earlier calculations \cite{GP,NWEG} illustrated 
this reporting different expressions for $Z_g$ depending on the vertex 
considered.} 
Since various mass scales contribute even in case power corrections can be 
disregarded, to treat $a_s$ including mass effects is also somewhat 
non practical. We treat the corresponding mass effects explicitly, which is outlined
in Section~\ref{secWF} below. 

\subsection{Mass Renormalization}

\vspace{1mm} \noindent
We choose the on--mass--shell scheme for quarks. In case of the heavy quarks 
the bare mass $\hat{m}$ is related to the renormalized mass by
\begin{eqnarray}
\hat{m}    &=& m + \hat{a}_s \delta m_1 + \hat{a}^2_s \delta m_2 + O(a_s^3) \\
\label{eqM1}
\delta m_1 &=& C_F S_\ep m \left(\frac{m^2}{\mu^2}\right)^{\ep/2} 
\left[ \frac{6}{\ep} - 4  + \left(4 + \frac{3}{4} \zeta_2\right) \ep \right] 
\\
\label{eqM2}
\delta m_2 &=& C_F S_\ep^2 m \left(\frac{{m}^2}{\mu^2}\right)^{\ep}
\Biggl[\frac{1}{\ep^2}\left(18 C_F + 22 C_A - 8 T_F (N_l+N_h)\right)
\nonumber\\ & &
+ \frac{1}{\ep} \left(-\frac{45}{2} C_F + \frac{91}{2} C_A - 14 T_F 
(N_l+N_h)\right) 
\nonumber\\ & &
+ C_F\left(\frac{199}{8} - \frac{51}{2} \zeta_2 + 48 \ln(2) \zeta_2 -12 
\zeta_3 \right) 
+ C_A\left(-\frac{605}{8} + \frac{5}{2} \zeta_2 - 24 \ln(2) \zeta_2 + 6 
\zeta_3 \right) \nonumber\\ && 
+ T_F \left[ N_l \left(\frac{45}{2} + 10 \zeta_2 \right) + N_h 
\left(\frac{69}{2} - 14 \zeta_2\right)\right]\Biggr]
\end{eqnarray}
(\ref{eqM1}) is easily obtained. The pole terms to (\ref{eqM2}) were given in
\cite{MASS1}, after charge renormalization, and the constant term in 
\cite{MASS2}, see also \cite{MASS3}.
The 3--loop corrections were given in \cite{MASS4}.
The renormalized mass is obtained absorbing $Z_m$ into the 
bare mass $\hat{m}$. Heavy quark mass effects occur also for massless quark
self-energies, see Section~\ref{secWF}.

\subsection{Operator Renormalization}

\vspace{1mm} \noindent
The local operators which emerge in the light cone expansion contain 
ultraviolet divergences. These are renormalized by the following $Z$--factors 
for the flavor non-singlet (NS), singlet (S), and pure singlet 
(PS) contributions. The formulae are partly generic and have to be adapted, 
e.g. for the three flavor non--singlet contributions. Here we suppress the 
argument $N$ in the anomalous dimensions $\gamma_{ij,k}$.
\begin{eqnarray}
Z_{\rm NS}(N,a_s,\ep) &=& 1 + a_s S_\ep \frac{\gamma_{\rm NS,0}}{\ep}
           + a_s^2 S_\ep^2 \left[\frac{1}{\ep^2}\left(\frac{1}{2} 
\gamma_{NS,0}^2 + \beta_0 \gamma_{\rm NS,0} \right) + \frac{1}{2 \ep} 
\gamma_{\rm NS,1} \right] 
\nonumber\\ 
& & + a_s^3 S_\ep^3\left[ \frac{1}{\ep^3} \left(\frac{1}{6} \gamma_{\rm 
NS,0}^3
+ \beta_0 \gamma_{\rm NS,0}^2 + \frac{4}{3} \beta_0^2 \gamma_{\rm NS,0}
\right) \right. \nonumber\\ & & \left.
+ \frac{1}{\ep^2}\left(
\frac{1}{2} \gamma_{\rm NS,0} \gamma_{\rm NS,1}
+\frac{2}{3} \beta_0 \gamma_{\rm NS,1}
+\frac{2}{3} \beta_1 \gamma_{\rm NS,0} \right)
+ \frac{1}{3 \ep} \gamma_{\rm NS,2} \right] 
\nonumber\\
\\
Z_{qq}(N,a_s,\ep) &=&
1 + a_s S_\ep \frac{\gamma_{qq,0}}{\ep}
  + a_s^{2} S_\ep^2 \left\{
  \frac{1}{\ep^2} \left[
  \frac{1}{2} \left(\gamma_{qq,0}^2 + \gamma_{qg,0}
\gamma_{gq,0}\right)   
+ \beta_0 \gamma_{qq,0}\right] 
+ \frac{1}{2 \ep} \gamma_{qq,1}
\right\}
\nonumber\\ & & 
+ a_s^{3} S_\ep^3 \left\{ \frac{1}{\ep^3} \left[
\frac{1}{6} \left(  \gamma_{qq,0}^3 
                 +2 \gamma_{qq,0} \gamma_{qg,0} \gamma_{gq,0}
                 +  \gamma_{qg,0} \gamma_{gg,0} \gamma_{gq,0}\right) 
+ \beta_0 \left(\gamma_{qq,0}^2 + \gamma_{qg,0} \gamma_{gq,0} \right)
\right. \right. \nonumber \\ & &
\left.
+ \frac{4}{3} \beta_0^2 \gamma_{qq,0}
\right] 
+ \frac{1}{\ep^2} \left[
  \frac{1}{2} \gamma_{qq,0} \gamma_{qq,1}
+ \frac{1}{3} \gamma_{qg,0} \gamma_{gq,1}
+ \frac{1}{6} \gamma_{qg,1} \gamma_{gq,0}
+ \frac{2}{3} \left( \beta_0 \gamma_{qq,1} + \beta_1 \gamma_{qq,0} \right) \right]
\nonumber\\ & &
\left.
+ \frac{\gamma_{qq,2}}{3 \ep} 
\right\}
\\
Z_{qg}(N,a_s,\ep) &=&
 a_s S_\ep \frac{\gamma_{qg,0}}{\ep}
  + a_s^{2} S_\ep^2 \left\{
  \frac{1}{\ep^2} \left[
  \frac{1}{2} \left( \gamma_{qg,0} \gamma_{gg,0}   
                   + \gamma_{qq,0} \gamma_{qg,0}\right)   
+ \beta_0 \gamma_{qg,0}\right] 
+ \frac{1}{2 \ep} \gamma_{qg,1}
\right\}
\nonumber\\ & &
+ a_s^{3} S_\ep^3
\left\{
  \frac{1}{\ep^3} 
\left[
  \frac{1}{6} \left(\gamma_{qg,0} \gamma_{gg,0}^2 
                  + \gamma_{qq,0} \gamma_{qg,0} \gamma_{gg,0} 
                  + \gamma_{qg,0} \gamma_{gq,0} \gamma_{qg,0}\right) 
\right.
\right.
\nonumber\\ & & \left. \left.
+ \beta_0 \left(\gamma_{qg,0} \gamma_{gg,0} + \gamma_{qq,0} \gamma_{qg,0}\right)
+\frac{4}{3} \beta_0^2 \gamma_{qg,0}\right]
\right. 
\nonumber\\ & & \left.
+ \frac{1}{\ep^2} \left[
\frac{1}{6} \left(
  \gamma_{qg,1} \gamma_{gg,0}
+ \gamma_{qq,1} \gamma_{qg,0} 
+ 2 \gamma_{qq,0} \gamma_{qg,1} 
+ 2 \gamma_{qg,0} \gamma_{gg,1}\right) 
+ \frac{2}{3} \left(\beta_0 \gamma_{qg,1} + \beta_1 \gamma_{qg,0}\right)
\right]
\right. \nonumber\\  & & \left.
+ \frac{\gamma_{qg,2}}{3 \ep} \right\}
\\
Z_{gq}(N,a_s,\ep) &=&
a_s S_\ep \frac{\gamma_{gq,0}}{\ep}
  + a_s^{2} S_\ep^2 \left\{
  \frac{1}{\ep^2} \left[
  \frac{1}{2} \left( \gamma_{gq,0} \gamma_{qq,0}   
                   + \gamma_{gg,0} \gamma_{gq,0}\right)   
+ \beta_0 \gamma_{gq,0}\right] 
+ \frac{1}{2 \ep} \gamma_{gq,1}
\right\}
\nonumber\\ & &
+ a_s^{3} S_\ep^3
\left\{
  \frac{1}{\ep^3} 
\left[
  \frac{1}{6} \left(\gamma_{gq,0} \gamma_{qq,0}^2 
                  + \gamma_{gg,0} \gamma_{gq,0} \gamma_{qq,0} 
                  + \gamma_{gq,0} \gamma_{qg,0} \gamma_{gq,0}\right) 
\right.
\right.
\nonumber\\ & & \left. \left.
+ \beta_0 \left(\gamma_{gq,0} \gamma_{qq,0} + \gamma_{gg,0} \gamma_{gq,0}\right)
+\frac{4}{3} \beta_0^2 \gamma_{gq,0}\right]
\right. 
\nonumber\\ & & \left.
+ \frac{1}{\ep^2} \left[
\frac{1}{6} \left(
  \gamma_{gq,1} \gamma_{qq,0}
+ \gamma_{gg,1} \gamma_{gq,0} 
+ 2 \gamma_{gg,0} \gamma_{gq,1} 
+ 2 \gamma_{gq,0} \gamma_{qq,1}\right) 
+ \frac{2}{3} \left(\beta_0 \gamma_{gq,1} + \beta_1 \gamma_{gq,0}\right)
\right]
\right. \nonumber\\  & & \left.
+ \frac{\gamma_{gq,2}}{3 \ep} \right\}
\end{eqnarray}\begin{eqnarray}
Z_{gg}(N,a_s,\ep) &=&
1 + a_s S_\ep \frac{\gamma_{gg,0}}{\ep}
  + a_s^{2} S_\ep^2 \left\{
  \frac{1}{\ep^2} \left[
  \frac{1}{2} \left(\gamma_{gg,0}^2 + \gamma_{gq,0}
\gamma_{qg,0}\right)   
+ \beta_0 \gamma_{gg,0}\right] 
+ \frac{1}{2 \ep} \gamma_{gg,1}
\right\}
\nonumber
\\ 
& & 
+ a_s^{3} S_\ep^3 \left\{ \frac{1}{\ep^3} \left[
\frac{1}{6} \left(  \gamma_{gg,0}^3 
                 +2 \gamma_{gg,0} \gamma_{gq,0} \gamma_{qg,0}
                 +  \gamma_{gq,0} \gamma_{qg,0} \gamma_{gg,0}\right) 
+ \beta_0 \left(\gamma_{gg,0}^2 + \gamma_{gq,0} \gamma_{qg,0} \right)
\right. \right. \nonumber 
\\ 
& &
  \left.
+ \frac{4}{3} \beta_0^2 \gamma_{gg,0}
\right]
+ \frac{1}{\ep^2} \left[
  \frac{1}{2} \gamma_{gg,0} \gamma_{gg,1}
+ \frac{1}{3} \gamma_{gq,0} \gamma_{qg,1}
+ \frac{1}{6} \gamma_{gq,1} \gamma_{qg,0}
+ \frac{2}{3} \left( \beta_0 \gamma_{gg,1} + \beta_1 \gamma_{gg,0} \right) \right]
\nonumber
\\ 
& &
\left.
+ \frac{\gamma_{gg,2}}{3 \ep} 
\right\}
\end{eqnarray}
The pure--singlet operator has the following $Z$--factor.
\begin{eqnarray}
Z_{qq}^{\rm PS}(N,a_s,\ep) &=&
  a_s^{2} S_\ep^2 \left[
  \frac{1}{2\ep^2} \gamma_{qg,0}\gamma_{gq,0}   
+ \frac{1}{2 \ep} \gamma_{qq,1}^{\rm PS} \right]
\nonumber\\ & & 
+ a_s^{3} S_\ep^3 \left[ \frac{1}{6 \ep^3} 
\left(
                 2 \gamma_{qq,0} \gamma_{qg,0} \gamma_{gq,0}
                 +  \gamma_{qg,0} \gamma_{gg,0} \gamma_{gq,0}\right) 
+ \frac{1}{6\ep^2} \left(
  2 \gamma_{qg,0} \gamma_{gq,1}
+ \gamma_{qg,1} \gamma_{gq,0}\right) \right.
\nonumber\\ & &
\left.
+ \frac{\gamma_{qq,2}^{\rm PS}}{3 \ep} 
\right]
\end{eqnarray}

The anomalous dimensions $\gamma_{ij,k}(N)$ are related to the splitting 
functions by
\begin{eqnarray}
\gamma_{ij,k}(N) = - \int_0^1 dz z^{N-1} P_{ij}^{(k)}(z)~.
\end{eqnarray}
The renormalized operators are obtained absorbing $Z_{\rm NS}, Z_{ij,\rm S}$, 
and $Z_{qq}^{\rm PS}$, 
into the bare operators, resp. operator matrix elements. 

\subsection{Wave Function Renormalization}
\label{secWF}

\vspace{1mm} \noindent
The external legs of the operator matrix elements are treated on--shell to be 
able to apply their factorization from the nucleon wave-functions in the light 
cone expansion as outlined above. Here the mass scale is set by a heavy 
quark mass. To the operator matrix elements also one-particle reducible diagrams 
contribute. If either the self-energy insertion on the external legs  
or the remainder diagram contain only massless lines, with the exception of 
the tree-level terms, the diagrams are vanishing since one of the factors has 
no scale. I.e. finite contributions are due to the self-energy insertions 
containing a massive line. The corresponding corrections are due to the 
massive contributions to the massless quark self--energy up to 3--loop order and 
the gluon self--energies up to 2--loop order. The former terms emerge in case of 
the flavor non--singlet terms $A_{qq,Q}^{\rm NS^l,(3)}$ and the latter in
$A_{Qg}^{(3)}$, while the pure singlet contributions $A_{Qq}^{\rm PS,(3)}$ 
obtain no corrections.
\subsubsection{Massless External Quark Lines}

\vspace{1mm} \noindent
The 2-loop correction reads 
\begin{eqnarray}
\Sigma^{(2)}_{ij} = a_s^2 \delta_{ij} T_F C_F 
\left(\frac{m^2}{\mu^2}\right)^\ep S_\ep^2
\left\{\frac{2}{\ep}  + \frac{5}{6}
+ \left[ \frac{89}{72} + 
\frac{\zeta_2}{2} \right] \ep + O(\ep^2) \right\} \cdot i p\!\!/~. 
\end{eqnarray}
At $O(a_s^2)$ this contribution implies, that the 1st moment of the 
non--singlet operator matrix element vanishes.
\subsubsection{External Gluon Lines}

\vspace{1mm} \noindent
The gluon vacuum polarization is given by
\begin{eqnarray}
\Pi_{\mu\nu}^{ab}(q) &=& \left[-g_{\mu\nu}q^2 +q_\mu q_\nu\right] 
\Pi^{ab}(q^2)~, 
\end{eqnarray}
with $a$ and $b$ the color indices. The 1-loop and 2-loop corrections read 
\begin{eqnarray}
\Pi_{(1)}^{ab}(0) &=& -i a_s \delta^{ab} T_F 
S_\ep \left(\frac{\hat{m}^2}{\mu^2}\right)^{\ep/2} \frac{4}{3} \left\{\frac{2}{\ep}  
+ 
\frac{\ep}{4} \zeta_2 + O(\ep^2)
\right\}\\ 
\Pi_{(2)}^{ab}(0) &=& -i a_s^2 \delta^{ab} T_F S_\ep^2 
\left(\frac{\hat{m}^2}{\mu^2}\right)^\ep \Biggl\{
C_F \left[ \frac{12}{\ep} + \frac{13}{3} + \left( \frac{35}{12} + 3 
\zeta_2\right)\ep \right] \nonumber\\ & &
+ C_A \left[ \frac{4}{\ep^2} - \frac{5}{\ep} 
-\left(\frac{13}{12} - \zeta_2\right) - \left(\frac{169}{144} + \frac{5}{4} 
\zeta_2 - \frac{\zeta_3}{3} \right) \ep\right] + O(\ep^2) \Biggr\}
\end{eqnarray}
The $C_F$-term can be compared with a corresponding contribution in the photon 
propagator, (\ref{eqrPI2}), before mass renormalization.
At 2--loop order the diagrams $u$ and $v$ from \cite{BUZA} and the term 
$Z^{-1,(1)}_{qg} \hat{A}_{gg}^{(1)}$ combine to
\begin{eqnarray}
\label{eqWF}
\left. \hat{A}_{Qg}^{(2)}\right|_{u,v} + Z^{-1,(1)}_{qg} \hat{A}_{gg}^{(1)}
&=& -2 \overline{a}_{Qg}^{(1)} \sum_{H=4}^6 \beta_{0,H} 
\left(\frac{m_H^2}{\mu^2} \right)^{\ep/2} \left(1 + \frac{\ep^2}{8} 
\zeta_2\right)\nonumber\\ 
&=& T_F^2 \frac{\zeta_2}{3}
P_{qg}^{(0)}(N) \sum_{H=4}^6 \left(\frac{m_H^2}{\mu^2} \right)^{\ep/2} \left(1 
+ 
\frac{\ep^2}{8} \zeta_2\right)~,  
\end{eqnarray}
with  
\begin{eqnarray}
\beta_{0,H} = - \frac{4}{3} T_F~.
\end{eqnarray}
(\ref{eqWF}) yields a finite contribution $\propto T_F^2$ in the 
$\overline{\rm MS}$ scheme. Our treatment differs from that in 
Ref.~\cite{BUZA} as we do not include the mass effects of (\ref{eqWF}) into
the running coupling, because we have chosen to define it in the 
$\overline{\rm MS}$--scheme. This is convenient for direct comparisons of 
the parton densities and the QCD--scale $\Lambda_{\rm QCD}$ measured in other 
analyzes of hard scattering cross sections.
\subsection{Mass Factorization}

\vspace{1mm} \noindent
The mass singularities are factored into the functions $\Gamma_{\rm NS}$,
$\Gamma_{ij,\rm S}$  and $\Gamma_{qq,\rm PS}$, respectively.
If all quarks were massless these functions were given by
\begin{eqnarray}
\Gamma_{\rm NS}        &=& Z_{\rm NS}^{-1}    \\
\Gamma_{ij, \rm S}     &=& Z_{ij, \rm S}^{-1} \\ 
\Gamma_{qq, \rm PS}    &=& Z_{qq, \rm PS}^{-1}~,
\end{eqnarray}
with
\begin{eqnarray}
\label{eqG1}
\Gamma_{\rm NS}(N,a_s,\ep) &=& 1 - a_s S_\ep \frac{\gamma_{\rm NS,0}}{\ep}
+ a_s^2 S_\ep^2 \left[\frac{1}{\ep^2}\left(\frac{1}{2} 
\gamma_{NS,0}^2 - \beta_0 \gamma_{\rm NS,0} \right) - \frac{1}{2 \ep} 
\gamma_{\rm NS,1} \right] 
\\
\Gamma_{ij,\rm S}(N, a_s, \ep) &=& \delta_{ij} 
- a_s S_\ep \frac{\gamma_{ij, \rm 0}}{\ep} 
           + a_s^2 S_\ep^2 \left[\frac{1}{\ep^2}\left(\frac{1}{2} 
\gamma_{ik, \rm 0} \gamma_{kj, \rm 0}
- \beta_0 \gamma_{ij,\rm 0} \right) - \frac{1}{2 \ep} 
\gamma_{ij, \rm 1} \right] 
\\
\label{eqG2}
\Gamma_{qq,\rm PS}(N,a_s,\ep) &=&  - a_s^2 S_\ep^2 
\left[\frac{1}{2\ep^2} 
\gamma_{qg,0} \gamma_{gq,0}  +\frac{1}{2 \ep} 
\gamma_{qq,\rm PS,1} \right]~. 
\end{eqnarray}
In the present calculation at least one quark line is massive in each diagram. 
Therefore the $\Gamma$--matrices (\ref{eqG1}--\ref{eqG2}) apply to the
parts of the diagrams which contain massless lines only, which are at most
2--loop sub-graphs. The mass factorization is therefore different in various 
sub--classes of contributing Feynman diagrams. 
The functions $\Gamma_{\rm NS}, \Gamma_{ij,\rm S}$, and $\Gamma_{qq,\rm PS}$  
do thus enter the 
renormalization of the operator matrix elements only in products with other
functions. The singularities contained in
$\Gamma_{\rm NS}, \Gamma_{ij,\rm S}$, and $\Gamma_{qq,\rm PS}$  are absorbed 
into the bare parton 
densities, which become scale--dependent in this way.
\subsection{The renormalized operator matrix elements}

\vspace{1mm}\noindent
The operator matrix element reads after charge and mass renormalization
\begin{eqnarray}
{\hat{A}}_{ij} &=& \delta_{ij} + a_s \hat{\hat{A}}_{ij}^{(1)} 
                  + a_s^2 \left[\hat{\hat{A}}^{(2)}_{ij} 
                  + \delta m_1 \frac{d}{dm} \hat{\hat{A}}_{ij}^{(1)}
                  + \delta a_s^{(1)} \hat{\hat{A}}_{ij}^{(1)}\right] \nonumber\\
               && + a_s^3 \left[
                  \hat{\hat{A}}^{(3)}_{ij} 
               + \delta m_1 \frac{d}{dm} \hat{\hat{A}}_{ij}^{(2)}
               + \delta m_2 \frac{d}{dm} \hat{\hat{A}}_{ij}^{(1)}  
               + \delta m_1^2 \frac{1}{2} \frac{d^2}{dm^2} 
                 \hat{\hat{A}}_{ij}^{(1)}  
+ \delta a_s^{(2)} \hat{\hat{A}}_{ij}^{(1)}
+ \delta a_s^{(1)} \hat{\hat{A}}_{ij}^{(2)}
\right]~.\nonumber\\
\end{eqnarray}
The renormalized operator matrix elements are obtained removing the ultraviolet singularities
and collinear singularities of the operator matrix elements,
\begin{eqnarray}
{A}_{ij} 
&=& Z^{-1}_{ik} {\hat{A}}_{kl} \Gamma^{-1}_{lj}  = \delta_{ij} 
+ a_s   A_{ij}^{(1)} 
+ a_s^2 A_{ij}^{(2)} 
+ a_s^3 A_{ij}^{(3)}~. 
\end{eqnarray}
Here self energy insertions containing massive lines in the external legs 
of the operator matrix elements have to be kept. 
\section{\boldmath The $O(\ep)$ Contributions}

\vspace{1mm}\noindent
The $O(\ep)$ contributions to $A_{Qg}, A_{Qq}^{\rm PS}$ and $A_{qq,Q}^{\rm NS}$
at $O(a_s^2)$ contribute to these quantities at $O(a_s^3)$ in 
combination with the various single pole terms emerging at 1--loop,
as outlined in Section~\ref{REN}. The diagrams to be evaluated are shown in 
\cite{BUZA,BBK1}. The results for the individual un-renormalized diagrams in $O(\ep)$ are given in 
Appendix~A.
As outlined before in Ref.~\cite{BBK1}, we calculate the
massive operator matrix elements performing the Feynman-parameter integrals directly,
i.e., without using the integration-by-parts method \cite{IBP} which was applied in
\cite{BUZA} up to the terms $O(\ep^0)$ before. We obtain representations in terms of
generalized hypergeometric functions \cite{HGF}, which may be expanded to the desired order in 
$\ep$. With increasing depth in $\ep$, more and more involved nested infinite sums 
are obtained, which depend on the Mellin--parameter $N$. These sums can be summed
applying analytic methods, as integral representations, and general summation
methods, as encoded in the  {\tt Sigma} package \cite{sigma}. 
We applied both methods to evaluate the sums which emerge at $O(\ep)$. 
The underlying algorithms of~{\tt Sigma} are based on a refined
version~\cite{RefinedDF} of Karr's difference field theory of
$\Pi\Sigma$-fields~\cite{Karr}. In this algebraic setting one can
represent completely algorithmically indefinite nested sums and
products without introducing any algebraic relations between them.
Note that this  general class of sum expressions covers as special cases,
e.g., the harmonic sums~\cite{BK1,SUMMER} or generalized nested harmonic
sums~cf.~\cite{GON,BBBL,PET,MUW}.
Given such an optimal representation, by introducing as less sums as 
possible,
various summation principles are available in~{\tt Sigma}. In this article we
applied the following strategy which has been generalized from the the 
hypergeometric
case~\cite{AequalB} to the $\Pi\Sigma$-field setting.
\begin{enumerate}
\item Given a definite sum that involves an extra parameter $N$,
for typical sums see the Appendix~B. We compute a recurrence relation
in $N$ that is fulfilled by 
the input sum. The underlying difference field
algorithms exploit Zeilberger's creative telescoping principle~\cite{AequalB}.
\item Then we solve the derived recurrence in terms of the so-called
d'Alembertian solutions~\cite{AequalB}. Since this class covers the harmonic 
sums,
we find all solutions in terms of harmonic sums.
\item Taking the initial values of the original input sum, we can combine
the solutions found from step~2 in order to arrive at a closed form 
representation
in terms of harmonic sums.
\end{enumerate}
A detailed example for the sum~\eqref{Beta25} with all its computation steps
has been carried out in~\cite{BBKS1}. In Appendix~B we present the details 
for the calculation of a further example, Eq.~(\ref{DoubleSum1}), in 
section~(\ref{sec:samp}). 

The results for new sums contributing 
are listed in Appendix~B. In the calculation also more well-known sums are occurring
which were found before in \cite{BBK1} or can be easily  solved using the {\tt FORM}--code 
\cite{FORM} {\tt summer} \cite{SUMMER}.

The $O(\ep)$ contribution to $A_{Qg}^{(2)}$ reads~:
%
%
\begin{eqnarray}
 \overline{a}_{Qg}^{(2)}(N)&=&
         T_FC_F\Biggl\{
                      \frac{N^2+N+2}
                           {N(N+1)(N+2)}
                   \Bigl(
                     16S_{2,1,1}
                    -8S_{3,1}
                    -8S_{2,1}S_1
                    +3S_4
                    -\frac{4}{3}S_3S_1
                    -\frac{1}{2}S^2_2
                    -S_2S^2_1\N\\
&&                  -\frac{1}{6}S^4_1
                    +2\zeta_2S_2
                    -2\zeta_2S^2_1
                    -\frac{8}{3}\zeta_3S_1
                   \Bigr)
                -8\frac{N^2-3N-2}
                       {N^2(N+1)(N+2)}S_{2,1}
                +\frac{2}{3}\frac{3N+2}
                         {N^2(N+2)}S^3_1\N\\
&&              +\frac{2}{3}\frac{3N^4+48N^3+43N^2-22N-8}
                         {N^2(N+1)^2(N+2)}S_3
                +2\frac{3N+2}
                       {N^2(N+2)}S_2S_1
                +4\frac{S_1}
                       {N^2}\zeta_2\N\\
&&              +\frac{2}{3}\frac{(N^2+N+2)(3N^2+3N+2)}
                                 {N^2(N+1)^2(N+2)}\zeta_3
                +\frac{P_{1}}
                                 {N^3(N+1)^3(N+2)}S_2\N\\
&&              +\frac{N^4-5N^3-32N^2-18N-4}
                                 {N^2(N+1)^2(N+2)}S^2_1
-\frac{5N^6+15N^5+36N^4+51N^3+25N^2+8N+4}
                                 {N^3(N+1)^3(N+2)}\zeta_2\N\\
&&              -2\frac{2N^5-2N^4-11N^3-19N^2-44N-12}
                                 {N^2(N+1)^3(N+2)}S_1
                -\frac{P_{2}}
                      {N^5(N+1)^5(N+2)}
                          \Biggr\}\N\\
&&       +T_F{C_A}\Biggl\{ 
                   \frac{N^2+N+2}
                         {N(N+1)(N+2)}
\nonumber\\ & &                      
\times \Bigl(
                        16S_{-2,1,1}
                       -4S_{2,1,1}
                       -8S_{-3,1}
                       -8S_{-2,2}
                       -4S_{3,1}
                       -\frac{2}{3}\beta'''
                       +9S_4\N\\
&&                     -16S_{-2,1}S_1
                       +\frac{40}{3}S_1S_3
                       +4\beta''S_1
                       -8\beta'S_2
                       +\frac{1}{2}S^2_2
                       -8\beta'S^2_1
                       +5S^2_1S_2
                       +\frac{1}{6}S^4_1
                       -\frac{10}{3}S_1\zeta_3
\N\\
&&                         -2S_2\zeta_2
                   -2S^2_1\zeta_2
                       -4\beta'\zeta_2
                       -\frac{17}{5}\zeta_2^2
                     \Bigr)
                  +\frac{4(N^2-N-4)}
                         {(N+1)^2(N+2)^2}
                      \Bigl(
                       -4S_{-2,1}
                       +\beta''
                       -4\beta'S_1
                     \Bigr)\N \\
&&                 -\frac{16}{3}\frac{N^5+10N^4+9N^3+3N^2+7N+6}
                             {(N-1)N^2(N+1)^2(N+2)^2}S_3
                   +2\frac{3N^3-12N^2-27N-2}
                          {N(N+1)^2(N+2)^2}S_2S_1\N\\
&&                 -\frac{2}{3}\frac{N^3+8N^2+11N+2}
                            {N(N+1)^2(N+2)^2}S^3_1
                   -8\frac{N^2+N-1}
                          {(N+1)^2(N+2)^2}\zeta_2S_1\N\\
&&                 -\frac{2}{3}\frac{9N^5-10N^4-11N^3+68N^2+24N+16}
                            {(N-1)N^2(N+1)^2(N+2)^2}\zeta_3
                   +8\frac{N^4+2N^3+7N^2+22N+20}
                          {(N+1)^3(N+2)^3}\beta'\N
\end{eqnarray}\begin{eqnarray}
&&                 -\frac{P_3}
                         {(N-1)N^3(N+1)^3(N+2)^3}S_2
                   -\frac{2P_{4}}
                          {(N-1)N^3(N+1)^3(N+2)^2}\zeta_2\N
\end{eqnarray}\begin{eqnarray}
&&                 -\frac{P_5}
                         {N(N+1)^3(N+2)^3}S^2_1
                   +\frac{2P_{6}}
                          {N(N+1)^4(N+2)^4}S_1
\N\\ & &                   -\frac{2P_{7}}
                          {(N-1)N^5(N+1)^5(N+2)^5}
                \Biggr\}~.\label{aQg2bar}
\end{eqnarray}
Here the argument $N$ of the harmonic sums, and $(N+1)$ in the function 
\begin{eqnarray}
\beta(N) &=& \frac{1}{2} \left[\psi\left(\frac{N+1}{2}\right) - \psi\left(\frac{N}{2}\right)\right] \\
S_{-1}(N) &=& (-1)^{N} \beta(N+1) - \ln(2)
\end{eqnarray}
and in the polynomials $P_i(N)$ was omitted as well as the factor
\begin{eqnarray}
\label{eqFA}
 S_{\ep}^2a_s^2\Bigl(\frac{m^2}{\mu^2}\Bigr)^{\ep}~.
\end{eqnarray}
In the gluon and  pure--singlet case
we did not write the overall factor
\begin{eqnarray}
 \frac{1+(-1)^N}{2}~.\N
\end{eqnarray}
It does not emerge generically in the non--singlet case. 
In accordance with the light--cone expansion only even integer moments
contribute in the present case.
The polynomials in Eq. (\ref{aQg2bar})
are 
\begin{eqnarray}
P_{1}&=&3N^6+30N^5+15N^4-64N^3-56N^2-20N-8~,\\
P_{2}&=&24N^{10}+136N^9+395N^8+704N^7+739N^6
         +407N^5+87N^4+27N^3+45N^2+24N+4~,
\nonumber\\
P_3&=&N^9+21N^8+85N^7+105N^6+42N^5+290N^4+600N^3+456N^2+256N+64\\
P_{4}&=&(N^3+3N^2+12N+4)(N^5-N^4+5N^2+N+2)~,\\
P_5&=&N^6+6N^5+7N^4+4N^3+18N^2+16N-8~,\\
P_{6}&=&2N^8+22N^7+117N^6+386N^5+759N^4+810N^3+396N^2+72N+32~,\\
P_{7}&=&4N^{15}+50N^{14}+267N^{13}+765N^{12}+1183N^{11}+682N^{10}
            -826N^9-1858N^8\\
         &&-1116N^7+457N^6+1500N^5+2268N^4+2400N^3
            +1392N^2+448N+64~. 
\end{eqnarray}
The flavor non-singlet and pure-singlet contributions read~:
\begin{eqnarray}
 \overline{a}_{qq,Q}^{{\rm NS},(2)}&=&
                      T_FC_F\Biggl\{
                        \frac{4}{3}S_4
                        +\frac{4}{3}S_2\zeta_2
                        -\frac{8}{9}S_1\zeta_3
                        -\frac{20}{9}S_3
                        -\frac{20}{9}S_1\zeta_2
                        +2\frac{3N^2+3N+2}
                                 {9N(N+1)}\zeta_3
                        +\frac{112}{27}S_2\N\\
&&                      +\frac{3N^4+6N^3+47N^2+20N-12}
                              {18N^2(N+1)^2}\zeta_2
                        -\frac{656}{81}S_1
                        +\frac{P_{8}}{648N^4(N+1)^4}
                               \Biggr\}
                          \label{aNS2qqQep}~. \\
P_{8}&=&1551N^8+6204N^7+15338N^6+17868N^5+8319N^4\N\\
&&+944N^3+528N^2-144N-432~. 
\end{eqnarray}\begin{eqnarray}
%
%
  \overline{a}_{Qq}^{{\rm PS},(2)}&=&
   T_FC_F\Biggl\{
           -2\frac{(5N^3+7N^2+4N+4)(N^2+5N+2)}
                  {(N-1)N^3(N+1)^3(N+2)^2}\Bigl(2S_2+\zeta_2\Bigr)\N\\
&&         -\frac{4}{3}\frac{(N^2+N+2)^2}
                 {(N-1)N^2(N+1)^2(N+2)}\Bigl(3S_3+\zeta_3\Bigr)
           +2\frac{P_{9}}
                  {(N-1)N^5(N+1)^5(N+2)^4}
       \Biggr\}.  \label{aPS2Qqep}
\N\\
\\
P_{9}&=&5N^{11}+62N^{10}+252N^9+374N^8-400N^6+38N^7-473N^5\N\\
&&           -682N^4-904N^3-592N^2-208N-32~.
\end{eqnarray}
The harmonic sums contributing to the individual diagrams, see Appendix~A, are 
listed in Table~1.
     \begin{table}[htb]
      \caption{\sf Complexity of the results in Mellin space 
      for the individual diagrams in the unpolarized case, 
      cf. \cite{BBK1}, up to $O(\ep)$}
      \label{table:results5}
      \begin{center}
       \renewcommand{\arraystretch}{1.1}
       \begin{tabular}{||l|c|c|c|c|c|c|c|c|c|c|c|c|c|r||}
        \hline \hline
  Diagram & $S_1$     & $S_2$        & $S_3$      & $S_4$       & $S_{-2}$ &
            $S_{-3}$  & $S_{-4}$     & $S_{2,1}$  & $S_{-2,1}$  & $S_{-2,2}$ &
            $S_{3,1}$ & $S_{-3,1}$   & $S_{2,1,1}$& $S_{-2,1,1}$ \\
        \hline \hline
          A             & &+&+& & & & & & & & & & &  \\
          B             &+&+&+&+& & & &+& & &+& &+&  \\
          C             & &+&+& & & & & & & & & & &  \\
          D             &+&+&+& & & & &+& & & & & &  \\
          E             &+&+&+& & & & &+& & & & & &  \\
          F             &+&+&+&+& & & &+& & & & &+&  \\
          G             &+&+&+& & & & &+& & & & & &  \\
          H             &+&+&+& & & & &+& & & & & &  \\
          I             &+&+&+&+&+&+&+&+&+&+&+&+&+& +\\
          J             & &+&+& & & & & & & & & & &  \\
          K             & &+&+& & & & & & & & & & &  \\
          L             &+&+&+&+& & & &+& & &+& &+&  \\
          M             & &+&+& & & & & & & & & & &  \\
          N             &+&+&+&+&+&+&+&+&+&+&+&+&+&+ \\
          O             &+&+&+&+& & & &+& & &+& &+&  \\
          P             &+&+&+&+& & & &+& & &+& &+&  \\
            \hline
          S             & &+&+& & & & & & & & & & &\\
          T             & &+&+& & & & & & & & & & &\\
            \hline
          ${\rm PS}_a$  & &+&+& & & & & & & & & & &\\
          ${\rm PS}_b$  & &+&+& & & & & & & & & & &\\
          \hline
          ${\rm NS}_a$  & & & & & & & & & & & & & &\\
          ${\rm NS}_b$  &+&+&+&+& & & & & & & & & &\\
          \hline\hline
       \end{tabular}
       \renewcommand{\arraystretch}{1.0}
      \end{center}
     \end{table}

Here we have already made use of the algebraic relations \cite{ALGEBRA}. Moreover,
two of the sums, $S_{-2,2}(N)$ and $S_{3,1}(N)$, can be related by structural 
relations \cite{STRUCT} to other harmonic sums, i.e., they lie in corresponding
equivalence classes and may be obtained by either rational argument relations 
and/or differentiation w.r.t. $N$. Reference to these equivalence classes is 
useful since the representation of these sums for $N~\epsilon~{\bf C}$ needs 
not to be derived newly, except of differentiation which is easily carried 
out. Therefore the two--loop massive operator matrix elements to $O(\epsi)$ depend 
on six basic harmonic sums.
\begin{figure}[h]
\begin{center}
\includegraphics[angle=0, width=13.0cm]{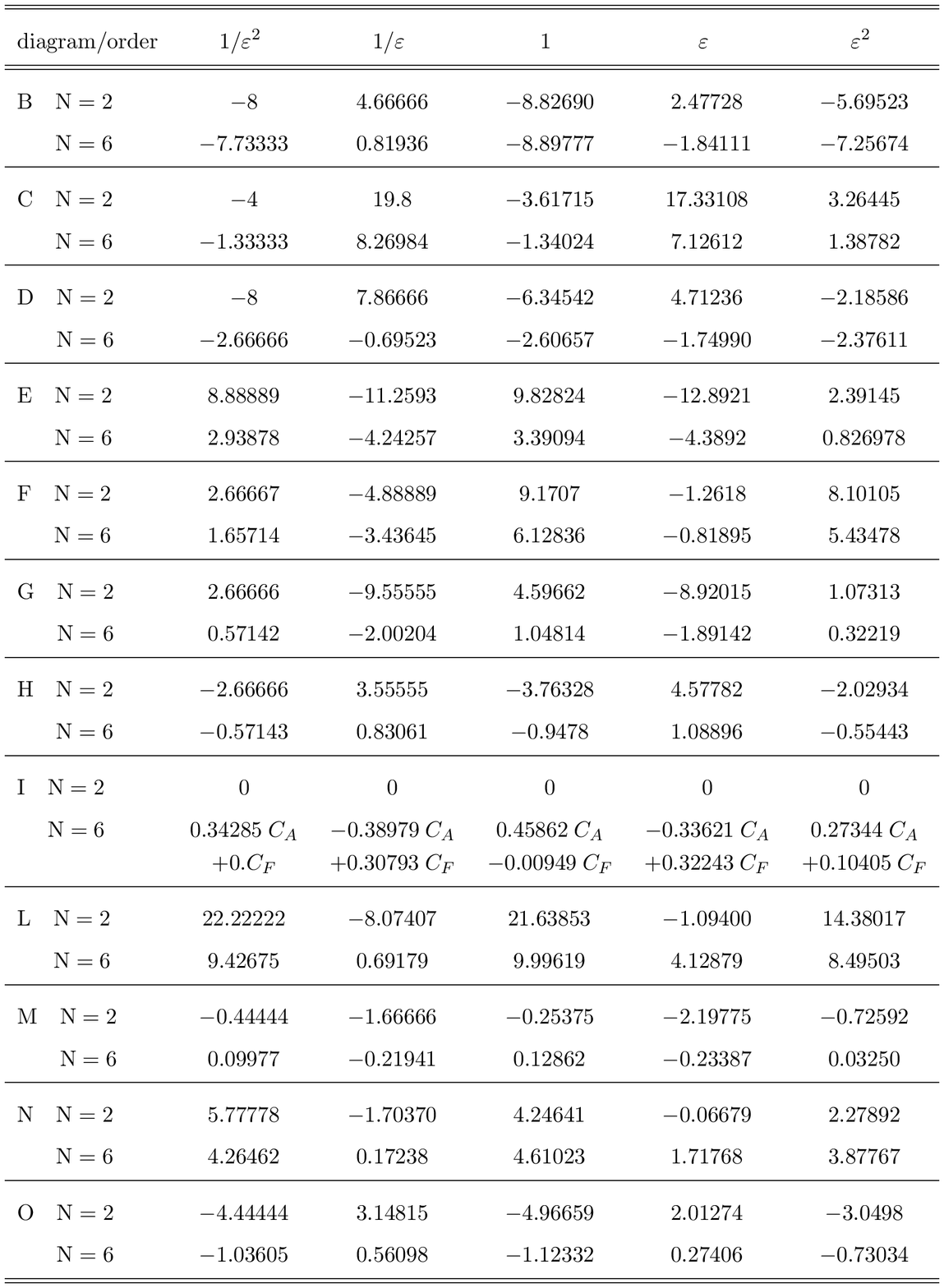} 
\end{center}
\end{figure}

\vspace{2mm}\noindent
{Table~2: \sf ~Numerical values for the moments $N =2,6$ for the 
the expansion of the un-renormalized matrix element $A_{Qg}^{2}$ for the terms
$O(1/\ep^2)$ to $O(\ep^2)$ for individual diagrams.} 

\vspace{2mm}

In the complex Mellin-$N$ plane these functions, up to more simple terms 
due to the soft- and virtual corrections, are meromorphic functions with 
poles at the non--positive integers, which possess both an analytic regular 
asymptotic representation and recursion relations, through which they may 
be calculated. To apply the results obtained in the present calculation in 
Mellin space in a QCD--analysis of deeply inelastic structure functions, their
scale evolution is first evaluated in Mellin space, incorporating
the heavy flavor Wilson coefficients for complex values of $N$. This requires
analytic continuations of the corresponding harmonic sums as worked out in \cite{ANCONT}.
The result in $x$--space is obtained by a single, fast numeric Mellin-inversion performed 
by a contour integral around the singularities of the problem located at the real axis left
to some value $r$. In case one wants to include small--$x$ resummations as 
well, this can be 
done in a similar way, see \cite{BV1}.

We performed an independent check on our calculation evaluating fixed moments 
in $N$ for the un-renormalized diagrams using the Mellin-Barnes method \cite{MB1,MB2}.
Here we use an extension of a method developed for massless propagators in \cite{BW}
to massive on--shell operator matrix elements \cite{BBK2}. The Mellin-Barnes integrals 
are evaluated numerically using the package {\tt MB} \cite{MB}. In Table~2, we present the
moments $N=2$ and $N=6$ for the more difficult two--loop diagrams, cf. \cite{BBK1}, for
the $O(1/\ep^2)$ to the $O(\ep^2)$ terms.

A further test for the Abelian part of the first moment of the un-renormalized massive operator
matrix element $A_{Qg}^{(2)}$ after mass renormalization, i.e., the term $\propto T_F C_F$, can be 
performed after analytic continuation of even values of $N$. This term is related to a 
corresponding contribution of the on--shell photon polarization function as 
noted in \cite{BUZA}.
We apply this method to the $O(\ep)$ term in Appendix~C and find agreement.

\section{Conclusions}

\vspace{1mm}\noindent
We calculated the $O(\ep)$ contributions to the massive operator matrix 
elements at $O(a_s^2)$ which contribute to the heavy flavor Wilson coefficients 
in deeply inelastic scattering to the non power-suppressed contributions. In the 
renormalization of the heavy flavor Wilson coefficients to 3--loop order they 
contribute together with with the single pole terms at $O(a_s)$. These terms, 
and the $O(a_s^2 \ep)$ contributions to the operator matrix element $A_{gg}(N)$ 
to be published soon, form all but the constant terms of the 3--loop heavy flavor
unpolarized operator matrix elements needed to describe the 3--loop heavy 
flavor
Wilson coefficients, together with the known 3--loop massless Wilson coefficients 
\cite{WIL3}, in the region $Q^2 \gg m^2$. In the calculation, we made use of the 
representation of the Feynman--parameter integrals in terms of generalized hypergeometric 
functions in a direct calculation, without applying the integration-by-parts method. The 
$\ep$--expansion leads to new infinite sums which had to be solved by analytic and 
advanced algebraic methods. We checked our results for finite values of $N$, using the
Mellin-Barnes method, and for a series of diagrams by a second program. Here, the calculation 
can be extended to higher corrections in $\ep$. For $N=1$, one may compare in addition the 
terms $\propto T_F C_F$ in $\overline{a}_{Qg}^{(2)}$ with the corresponding contribution 
in the 2--loop on--shell photon propagator. The terms $\overline{a}^{(2)}(N)$ 
can
be expressed in terms of polynomials of the basic nested harmonic sums up to weight ${\sf w=4}$
and derivatives thereof. They belong to the complexity-class of the general two-loop 
Wilson coefficients or hard scattering cross sections in massless QED and QCD found for 
space-- and time--like unpolarized and polarized anomalous dimensions, massless Wilson coefficients 
for deeply inelastic scattering, parton fragmentation, the Drell--Yan process, hadronic Higgs-- and 
pseudoscalar Higgs production in the heavy mass limit as well as the soft- and virtual contribution 
to Bhabha-scattering, cf.~\cite{MATH}, and are described by six basic functions and their derivatives 
in Mellin space. Their analytic continuation to complex values of $N$ is known in explicit form. 
The package {\tt Sigma} \cite{sigma} proved to be a useful tool to solve the sums occurring in the 
present problem and was extended accordingly. One advantage to seek for solutions of the recurrences
emerging in $\Pi \Sigma$-fields consists in finding irreducible structures for the representation.
In the present calculation these were nested harmonic sums. In even more complicated single scale 
problems in higher orders, this needs not to be the case. The new basis elements, however, would be 
uniquely found applying the present procedure.

\newpage
\appendix
\section{\bf\boldmath 
The $O(\ep)$ Terms for the Individual Diagrams}

\vspace{1mm}\noindent
In the following we list the results for the individual diagrams for comparisons
and to illustrate the analytic structures emerging in the calculation.  
The calculation is performed in Feynman gauge. Again 
we suppress the factor (\ref{eqFA}) and the argument $N$ in the sums and 
polynomials.
\vspace*{-3mm}
\begin{eqnarray}
\overline{A}^{Qg}_a&=&T_FC_F\Biggl\{
           4\frac{3S_3+\zeta_3}{3N^2(N+1)}
          +2\frac{2N^3-N-2}{N^3(N+1)^2(N+2)}\Bigl(2S_2+\zeta_2\Bigr)
        -\frac{2\PO_1}
                {N^5(N+1)^4(N+2)^3}\Biggr\} ~,       \label{resAep} \\
\PO_1&=&2N^9-16N^8-89N^7-166N^6-135N^5-6N^4+85N^3+94N^2+44N+8~.\N\\
%
\overline{A}^{Qg}_b&=&T_FC_F\Biggl\{
           \frac{1}{N}\Bigl(
               8S_{2,1,1}
              -8S_{3,1}
              +11S_4
              -8S_{2,1}S_1
              -\frac{4}{3}S_3S_1
              +\frac{7}{2}S^2_2
              -S_2S^2_1
              -\frac{1}{6}S^4_1
              +6S_2\zeta_2\N\\
&&            -2S^2_1\zeta_2
              -\frac{8}{3}S_1\zeta_3
              +\frac{8}{3}\zeta_3
                       \Bigr)
          +\frac{N^2+7N+2}{N(N+1)(N+2)}\Bigl(
               8S_{2,1}
              +2S_2S_1
              +\frac{2}{3}S^3_1
              +4S_1\zeta_2
                                        \Bigr)\N\\
&&        -4\frac{N^5+3N^4+19N^3+37N^2+16N+4}{N^2(N+1)^2(N+2)^2}S_2
          -4\frac{N^3+9N^2+8N+4}{N^2(N+2)^2}S^2_1\N\\
&&        -8\frac{N^2+5N+2}{N(N+1)(N+2)}\zeta_2
          +\frac{4}{3}\frac{N^2-17N+2}{N(N+1)(N+2)}S_3
          +\frac{16\PO_2}{N^2(N+1)^3(N+2)^3}S_1\N\\
&&        -\frac{16\PO_3}{N(N+1)^4(N+2)^3}
                            \Biggr\} ~,       \label{resBep} \\
\PO_2&=&N^7+14N^6+65N^5+153N^4+197N^3+134N^2+44N+8~,\N\\
\PO_3&=&2N^7+27N^6+130N^5+306N^4+385N^3+266N^2+100N+16~.\N\\
\overline{A}^{Qg}_c&=&T_FC_F\Biggl\{
               \frac{2}{N}\Bigl(5S_3-\frac{1}{3}\zeta_3\Bigr)
              -2\frac{7N^3+29N^2+15N+2}
                     {N^2(N+1)(N+2)}S_2\N\\
&&            +\frac{13N^4+82N^3+82N^2+N-6}
                     {N^2(N+1)(N+2)(N+3)}\zeta_2
              +\frac{\PO_4}
                    {N^4(N+1)^3(N+2)^3(N+3)}
                            \Biggr\} ~,       \label{resCep} \\
\PO_4&=&32N^{10}+448N^9+2177N^8+5123N^7+6312N^6+3863N^5+902N^4\N\\
&&    +9N^3-74N^2-68N-24~.  \N\\
%
\overline{A}^{Qg}_d&=&T_FC_F\Biggl\{
           \frac{1}{N}\Bigl(
                  -4S_{2,1}
                  -\frac{2}{3}S_3
                  -S_2S_1
                  -\frac{1}{3}S^3_1
                  -2S_1\zeta_2
                  -\frac{4}{3}\zeta_3
                       \Bigr)\N\\
&&
          +\frac{N^4+8N^3+43N^2+36N+12}
                {N^2(N+1)^2(N+2)}\Bigl(S_2+S^2_1\Bigr)
          +2\frac{N^3+10N^2+59N+42}{N(N+1)(N+2)(N+3)}\zeta_2\N\\
&&        -\frac{4\PO_5}
                {N^2(N+1)^3(N+2)^2}S_1
          +\frac{4\PO_6}
                {N(N+1)^4(N+2)^3(N+3)}
                            \Biggr\} ~,       \label{resDep} \\
\PO_5&=&N^6+8N^5+79N^4+207N^3+205N^2+96N+20~, \N\\
\PO_6&=&2N^8+24N^7+262N^6+1371N^5+3514N^4+4775N^3+3544N^2+1404N+240~. \N\\
%
\overline{A}^{Qg}_e&=&T_F\Biggl[C_F-\frac{C_A}{2}\Biggr] \Biggl\{
      -2\frac{N+2}{N(N+1)}\Bigl(2S_{2,1}+S_1\zeta_2\Bigr)
      -\frac{2}{3}\frac{13N^4+60N^3+111N^2+4N-36}
               {N(N+1)^2(N+2)(N+3)}S_3\N\\
&&    -\frac{1}{3}\frac{N^3-N^2-8N-36}
            {N(N+1)(N+2)(N+3)}\Bigl(3S_2S_1+S^3_1\Bigr)
      +\frac{4}{3}\frac{N+3}
               {(N+1)^2}\zeta_3
      -2\frac{3N^3+9N^2+12N+4}
             {N(N+1)^3(N+2)}\zeta_2\N
\end{eqnarray}
\newpage
\begin{eqnarray}
&&    +\frac{\PO_{7}}
            {N^2(N+1)^3(N+2)(N+3)}S_2
      +\frac{4N^5+11N^4+15N^3-86N^2-92N-24}
            {N^2(N+1)^2(N+2)(N+3)}S^2_1\N
\\
&&    -2\frac{\PO_{8}}
             {N^2(N+1)^3(N+2)^2(N+3)}S_1
      -2\frac{\PO_{9}}
             {N^3(N+1)^5(N+2)^3(N+3)}
                            \Biggr\} ~,       \label{resEep} \\
\PO_{7}&=&20N^6+119N^5+290N^4+105N^3-290N^2-212N-24~, \N \\
\PO_{8}&=&8N^7+62N^6+181N^5+127N^4-226N^3-404N^2-296N-96~, \N \\
\PO_{9}&=&38N^{10}+394N^9+1775N^8+4358N^7+6323N^6+5788N^5+3626N^4+1462N^3\N\\
&&       +100N^2-184N-48~. \N \\
%
\overline{A}^{Qg}_f&=&T_F\Biggl[C_F-\frac{C_A}{2}\Biggr] \Biggl\{
                         \frac{4}{N}\Bigl(
                                  2S_{2,1,1}-2S_4
                                 -S^2_2-\zeta_2S_2
                                    \Bigr)
                         +\frac{16(S_1-1)\zeta_3}{3(N+1)(N+2)}\N\\
&&                       -8\frac{2N^2-5N-2}
                                 {N^2(N+1)(N+2)}S_{2,1}
                         +\frac{4}{3}\frac{26N^2-33N-10}
                                  {N^2(N+1)(N+2)}S_3
\nonumber\\ & &                         
+\frac{2}{3}\frac{2N^2-3N+2}
                                  {N^2(N+1)(N+2)}\Bigl(3S_2S_1+S^3_1\Bigr)
+4\frac{5N+2}
                                {N^2(N+1)(N+2)}S_1\zeta_2
\N\\
&&                       
                         -2\frac{2N^3+15N^2+12N-4}
                                {N(N+1)^2(N+2)^2}S^2_1
                         +2\frac{14N^3+85N^2+132N+52}
                                {N(N+1)^2(N+2)^2}S_2\N\\
&&                       +4\frac{2N^5+18N^4+25N^3+10N^2+44N+56}
                                {N(N+1)^3(N+2)^3}S_1
                         -\frac{8\zeta_2}
                               {(N+1)^2(N+2)}\N\\
&&                       -8\frac{6N^5+52N^4+178N^3+309N^2+264N+84}
                                 {(N+1)^4(N+2)^3}
                            \Biggr\} ~.       \label{resFep}\\
%
\overline{A}^{Qg}_g&=&T_FC_F \Biggl\{
                      \frac{ 12S_{2,1}-58S_3+3S_2S_1+S^3_1
                            +6\zeta_2S_1+8\zeta_3}
                           {3(N+1)(N+2)}
                       +\frac{(9N+8)(5N^2+9N-1)}
                             {N(N+1)^2(N+2)^2}S_2\N\\
&&                     -\frac{3N^3+31N^2+45N+8}
                             {N(N+1)^2(N+2)^2}S^2_1
                       -2\frac{17N^2+47N+28}
                              {(N+1)^2(N+2)^2}\zeta_2\N\\
&&                     +2\frac{6N^5+104N^4+376N^3+514N^2+277N+48}
                              {N(N+1)^3(N+2)^3}S_1
                       -\frac{2\PO_{10}}
                              {(N+1)^4(N+2)^4}
                            \Biggr\} ~,       \label{resGep} \\
\PO_{10}&=&74N^6+722N^5+2697N^4+4960N^3+4700N^2+2143N+368~. \N \\
\overline{A}^{Qg}_h&=&T_F\Biggl[C_F-\frac{C_A}{2}\Biggr]\Biggl\{
      \frac{4(N+3)}
             {N(N+1)(N+2)}\Bigl(2S_{2,1}+S_1\zeta_2\Bigr)
      -\frac{4}{3}\frac{N^2-26N+9}
                       {N (N+1)(N+2)(N+3)}S_3\N\\
&&    -\frac{2}{3}\frac{N^2-2N+9}
                       {N(N+1)(N+2)(N+3)}\Bigl(3S_2S_1+S^3_1\Bigr)
      -\frac{8\zeta_3}{3(N+1)(N+2)}
\N\\ & &
      -2\frac{N^2+7N+8}
             {(N+1)^2(N+2)^2}\zeta_2
-\frac{N^4+115N^3+335N^2+265N+84}
             {N(N+1)^2(N+2)^2(N+3)}S_2
\N\\ &&
      +\frac{7N^4+N^3-31N^2-37N-36}
             {N(N+1)^2(N+2)^2(N+3)}S^2_1\N\\
&&    -\frac{2\PO_{11}}
             {N(N+1)^3(N+2)^3(N+3)}S_1
      +\frac{2\PO_{12}}
             {(N+1)^4(N+2)^4(N+3)}
                            \Biggr\} ~,       \label{resHep} \\
\PO_{11}&=&10N^6+38N^5-43N^4-533N^3-1267N^2-1269N-456~, \N 
\end{eqnarray} \begin{eqnarray}
\PO_{12}&=&36N^7+344N^6+1287N^5+2143N^4+818N^3-2153N^2-2795N-960~. \N\\
\overline{A}^{Qg}_i&=&T_FC_A\Biggl\{
            \frac{8}{N+2}\Bigl(
                4S_{-2,1,1}
               -2S_{-3,1}
               -2S_{-2,2}
               -4S_{-2,1}S_1
               +S_{-4}
               +2S_{-3}S_1
\N\\ && 
              +2S_{-2}S_2
               +2S_{-2}S^2_1
+S_{-2}\zeta_2
                         \Bigr)
           +8\frac{N^2-N-4}{(N+1)(N+2)^2}\Bigl(
                2S_{-2,1}
               -S_{-3}
               -2S_{-2}S_1
                         \Bigr)
\N\\ &&
           +\frac{1}{(N+1)(N+2)}\Biggl(
               4(4N+7)S_{2,1,1} \N
-4(6N+7)S_{3,1}
               -4(4N+5)S_{2,1}S_1
               +\frac{32N+27}{2}S_4
\N\\ &&
               +\frac{2}{3}(36N+35)S_3S_1
               -\frac{16N+25}{4}S^2_2            +\frac{16N+15}{2}S_2S^2_1
               -\frac{1}{12}S^4_1 \N\\ &&
               -S^2_1\zeta_2
               +\frac{4}{3}S_1\zeta_3
               +(4N+3)S_2\zeta_2
                         \Biggr)
           +2\frac{N^3+9N^2+17N+8}
                  {N(N+1)^2(N+2)^2}S_1\zeta_2\N\\
&&         +4\frac{2N^4+N^3-N^2+9N+8}
                  {N(N+1)^2(N+2)^2}S_{2,1}
           -\frac{2}{3}\frac{18N^4+3N^3-67N^2-39N+8}
                  {N(N+1)^2(N+2)^2}S_3 \N\\
&&         -\frac{8N^4-3N^3-47N^2-29N+8}
                  {N(N+1)^2(N+2)^2}S_2S_1
           +\frac{1}{3}\frac{3N^3+7N^2-3N-8}
                  {N(N+1)^2(N+2)^2}S^3_1
\N\\ &&
           -\frac{4(N+4)\zeta_3}
                 {3(N+1)(N+2)^2}
         +8\frac{N^4+2N^3+7N^2+22N+20}
                  {(N+1)^2(N+2)^3}S_{-2}
           +\frac{\PO_{13}S_2}
                  {N(N+1)^3(N+2)^3}\N\\
&&         -\frac{6N^5+40N^4+92N^3+94N^2+49N+16}
                  {N(N+1)^3(N+2)^3}S^2_1
           -2\frac{2N^3+8N^2+19N+16}
                  {(N+1)^2(N+2)^3}\zeta_2\N\\
&&         +\frac{2\PO_{14}}
                  {N(N+1)^4(N+2)^4}S_1
           -\frac{2\PO_{15}}
                  {(N+1)^4(N+2)^5}
                            \Biggr\} \N\\
                   &&+T_FC_F\Biggl\{
            \frac{1}{(N+1)(N+2)}\Bigl(
                 -32S_{2,1,1}
                 +16S_{3,1}
                 -6S_4
                 +\frac{8}{3}S_3S_1
                 +16S_{2,1}S_1
\N\\ &&                 
+S^2_2
                 +2S_2S^2_1
                 +\frac{1}{3}S^4_1\N              -4S_2\zeta_2
                 +4S^2_1\zeta_2
                 +32\zeta_2
                 +224
                               \Bigr)
            -16\frac{2S_{2,1}+S_1\zeta_2}
                  {N(N+2)}
\N\\ &&
            -\frac{4}{3}\frac{3N-2}
                             {N(N+1)(N+2)}\Bigl(3S_2S_1+S^3_1\Bigr)
          +\frac{16}{3}\frac{3N+1}
                              {N(N+1)(N+2)}S_3
\N\\ &&
            -8\frac{2N^2+2N-1}
                   {N(N+1)^2(N+2)}S_2
            +8\frac{3N^2+3N+1}
                   {N(N+1)^2(N+2)}S^2_1\N\\
&&          -8\frac{11N^3+33N^2+38N+14}
                   {N(N+1)^3(N+2)}S_1
                            \Biggr\}~, \label{resIep}\\
\PO_{13}&=&4N^6+10N^5+12N^4+8N^3-18N^2-37N-16~,\N\\
\PO_{14}&=&10N^7+114N^6+533N^5+1374N^4+2144N^3+2027N^2+1057N+224~,\N\\
\PO_{15}&=&20N^7+236N^6+1202N^5+3384N^4+5688N^3+5720N^2+3195N+768~.\N\\
\overline{A}^{Qg}_j&=&T_FC_A\Biggl\{
         -\frac{2}{3}\frac{4N^2+4N-5}
                          {N^2(N+1)^2}\Bigl(3S_3+\zeta_3\Bigr)
         +\frac{\PO_{16}}
                {N^5(N+1)^5(N+2)^3}\N\\
&&       +\frac{4N^5+22N^4+11N^3+13N^2+35N+10} 
                {N^3(N+1)^3(N+2)}\Bigl(2S_2+\zeta_2\Bigr)
                            \Biggr\} ~,       \label{resJep} 
\end{eqnarray} \begin{eqnarray}
\PO_{16}&=&28N^{10}+148N^9+342N^8+285N^7-212N^6-114N^5+1117N^4+1587N^3+826N^2\N\\
      && +260N+40~.  \N\\
%
\overline{A}^{Qg}_k&=&T_FC_A\Biggl\{
        \frac{2}{3}\frac{3N^2-23N-20}
                        {(N-1)N(N+1)^2(N+2)}\Bigl(3S_3+\zeta_3\Bigr)
        -\frac{\PO_{17}}
              {(N-1)N(N+1)^5(N+2)^4}\N\\
&&     -\frac{10N^4+7N^3+51N^2+172N+112}
              {(N-1)N(N+1)^3(N+2)^2}\Bigl(2S_2+\zeta_2\Bigr)
                            \Biggr\} ~,       \label{resKep} \\
\PO_{17}&=&14N^8+70N^7+96N^6-375N^5-1493N^4-1056N^3+2392N^2+4192N+1792~.  \\
\overline{A}^{Qg}_l&=&T_FC_A\Biggl\{
               \frac{1}{N}\Bigl(
                -4S_{2,1,1}
                +4S_{3,1}
                +\frac{5}{2}S_4
                +4S_{2,1}S_1
                +\frac{2}{3}S_3S_1
                +\frac{9}{4}S^2_2
                +\frac{1}{2}S_2S^2_1
                +S_2\zeta_2
                +\frac{1}{12}S^4_1\N\\
&&              +S^2_1\zeta_2
                +\frac{4}{3}S_1\zeta_3
                          \Bigr)
              +\frac{2}{N(N+1)}\Bigl(
                -4S_{2,1}
                -S_2S_1
                -\frac{1}{3}S^3_1
                -2S_1\zeta_2
                          \Bigr) \N\\
&&            -\frac{2}{3}\frac{6N^3+5N^2-4N-6}
                               {N^2(N+1)^2}S_3
              +\frac{8N^5+18N^4+11N^3+N^2+6N+4}
                    {N^3(N+1)^3}S_2\N\\
&&            -\frac{(N+2)(2N+1)}{N^2(N+1)^2}S^2_1 
              +2\frac{7N^3+15N^2+7N+4}
                     {N^2(N+1)^3}S_1
              +\frac{2}{3}\frac{2N^3+5N^2+4N+2}
                               {N^2(N+1)^2}\zeta_3\N\\
&&            -\frac{\PO_{18}\zeta_2}
                    {N^3(N+1)^3(N+2)}
              -\frac{\PO_{19}}
                    {N^5(N+1)^5(N+2)}
                            \Biggr\} ~,       \label{resLep} \\
\PO_{18}&=&4N^6+30N^5+55N^4+38N^3+4N^2-10N-4~,\N\\
\PO_{19}&=&16N^{10}+152N^9+454N^8+628N^7+447N^6+180N^5+52N^4\N\\
 &&      -10N^3-28N^2-18N-4~,\N\\
\overline{A}^{Qg}_m&=&T_FC_A\Biggl\{
               \frac{2}{3}\frac{N^2-2N-2}
                               {N^2(N+1)^2}\Bigl(3S_3+\zeta_3\Bigr)
               -\frac{2N^5+11N^4+12N^3+2N^2+6N+4}
                      {N^3(N+1)^3(N+2)}\Bigl(2S_2+\zeta_2\Bigr)\N\\
&&             -\frac{6N^8+28N^7+53N^6+30N^5-14N^4+2N^3+18N^2+14N+4}
                      {N^5(N+1)^5(N+2)}
                            \Biggr\} ~,       \label{resMep} \\
\overline{A}^{Qg}_n&=&T_FC_A\Biggl\{
              \frac{4(N-1)}{N(N+1)}\Bigl(
              -4S_{-2,1,1}
              +2S_{-3,1}
              +2S_{-2,2}
              -S_{-4}
              +4S_{-2,1}S_1
              -2S_{-3}S_1
              -2S_{-2}S_2 \N\\
&&            -2S_{-2}S^2_1
              -S_{-2}\zeta_2
                               \Bigr)
            +\frac{2N^2+3N+2}{24N(N+1)(N+2)}\Bigl(
              S^4_1
              +12S^2_1\zeta_2
              +16\zeta_3S_1
                               \Bigr)\N\\
&&          -2\frac{6N^2+7N-6}
                   {N(N+1)(N+2)}\Bigl(S_{2,1,1}-S_{2,1}S_1\Bigr)
            +2\frac{8N^2+9N-10}
                   {N(N+1)(N+2)}S_{3,1}
            -\frac{54N^2+97N-10}
                  {4N(N+1)(N+2)}S_4\N\\
&&          -\frac{34N^2+33N-74}
                  {3N(N+1)(N+2)}S_3S_1
            +\frac{2N^2-37N-78}
                  {8N(N+1)(N+2)}S^2_2
            -\frac{14N^2+13N-34}
                  {4N(N+1)(N+2)}S_2S^2_1\N\\
&&          -\frac{10N^2+21N+6}
                  {2N(N+1)(N+2)}S_2\zeta_2
            +8\frac{N^2-N-4}{(N+1)^2(N+2)}\Bigl(
              -2S_{-2,1}
              +S_{-3}
              +2S_{-2}S_1
                               \Bigr)\N\\
&&          -2\frac{5N^5+10N^4-20N^3-62N^2-40N-8}
                   {N^2(N+1)^2(N+2)^2}S_{2,1}
            -\frac{2N(N+3)}
                   {3(N+1)^2(N+2)}\zeta_3\N\\
&&          +\frac{35N^5+66N^4-118N^3-294N^2-152N-40}
                             {3N^2(N+1)^2(N+2)^2}S_3
\N\\ &&          
  -\frac{N^5+6N^4+4N^3-30N^2-40N-8}
                  {N^2(N+1)^2(N+2)^2}\zeta_2S_1          
\N
\end{eqnarray}\begin{eqnarray}
&& +\frac{15N^5+10N^4-100N^3-98N^2+40N+8}
                  {2N^2(N+1)^2(N+2)^2}S_2S_1 \N\\ &&
            -\frac{N^5+6N^4+4N^3-30N^2-40N-8}
                  {6N^2(N+1)^2(N+2)^2}S^3_1
          -8\frac{N^4+2N^3+7N^2+22N+20}
                   {(N+1)^3(N+2)^2}S_{-2} \N\\ &&
            -\frac{\PO_{20}S_2-\PO_{21}S^2_1}
                  {2N(N+1)^3(N+2)^3}
         +\frac{2N^4+11N^3+15N^2+12N+8}
                  {(N+1)^3(N+2)^2}\zeta_2 \N\\ &&
            -\frac{\PO_{22}S_1}
                  {N(N+1)^4(N+2)^4}
            +\frac{\PO_{23}}
                  {(N+1)^5(N+2)^4}
                            \Biggr\} ~,       \label{resNep} \\
\PO_{20}&=&6N^6+36N^5+60N^4-99N^3-390N^2-316N-40~,\N\\
\PO_{21}&=&2N^6+20N^5+40N^4-45N^3-170N^2-100N+8~,\N
\\
\PO_{22}&=&4N^8+50N^7+224N^6+544N^5+927N^4+1140N^3+712N^2-64N-208~,\N\\
\PO_{23}&=&8N^8+86N^7+370N^6+805N^5+807N^4-16N^3-772N^2-568N-96~.\N\\
%
\overline{A}^{Qg}_o&=&T_FC_A\Biggl\{
              \frac{1}{2N(N+2)}\Bigl(
                   8S_{2,1,1}
                  -8S_{3,1}
                  -5S_4
                  -8S_1S_{2,1}
                  -\frac{4}{3}S_3S_1
                  -\frac{9}{2}S^2_2
                  -S_2S^2_1
\N\\
&&                
                  -\frac{1}{6}S^4_1
                  -2S_2\zeta_2
                  -2S^2_1\zeta_2
                  -\frac{8}{3}S_1\zeta_3
                               \Bigr)
             +\frac{2N^2+9N+12}{2N(N+1)(N+2)^2}\Bigl(
                   4S_{2,1}
                  +S_2S_1
                  +\frac{1}{3}S^3_1
                  +2S_1\zeta_2
                               \Bigr)
\N\\ &&                  
-\frac{2}{3}\frac{N^2+7N+8}
                                   {(N+1)^2(N+2)^2}\zeta_3\N\\
&&                +\frac{1}{3}\frac{14N^3+41N^2+51N+36}
                                   {N(N+1)^2(N+2)^2}S_3
                  -\frac{12N^5+124N^4+472N^3+817N^2+641N+176}
                        {2N(N+1)^3(N+2)^3}S_2\N\\
&&                +\frac{4N^4+16N^3-4N^2-61N-48}
                        {2N(N+1)^2(N+2)^3}S^2_1
                  +\frac{N(11N^3+56N^2+92N+49)}
                        {(N+1)^3(N+2)^3}\zeta_2\N\\
&&                -\frac{\PO_{24}S_1}
                         {N(N+1)^3(N+2)^4}
                  +\frac{\PO_{25}}
                         {(N+1)^5(N+2)^5}
                            \Biggr\} ~,       \label{resOep} \\
\PO_{24}&=&16N^6+101N^5+194N^4+4N^3-421N^2-501N-192~, \N\\
\PO_{25}&=&62N^8+668N^7+3073N^6+7849N^5+12052N^4+11127N^3+5640N^2+1065N
         -128~. \N\\
%
\overline{A}^{Qg}_p&=&T_FC_A\Biggl\{
              \frac{N-4}{2N(N+1)(N+2)}\Bigl(
                   4S_{2,1,1}
                  -4S_{3,1}
                  -4S_{2,1}S_1
                  -\frac{2}{3}S_3S_1
                  -\frac{1}{2}S_2S^2_1
                  -\frac{1}{12}S^4_1
                  -S^2_1\zeta_2\N\\
&&                -\frac{4}{3}S_1\zeta_3
                               \Bigr)
             +\frac{N^3-17N^2-41N-16}{2N(N+1)^2(N+2)^2}\Bigl(
                   4S_{2,1}
                  +S_2S_1
                  +\frac{1}{3}S^3_1
                  +2S_1\zeta_2
                               \Bigr)
\N\\ &&
             +\frac{11N+20}
                   {4N(N+1)(N+2)}S_4\N\\
&&           +\frac{7N+36}
                   {8N(N+1)(N+2)}S^2_2
             +\frac{3N+4}
                   {2N(N+1)(N+2)}S_2\zeta_2
             -\frac{1}{3}\frac{11N^3+17N^2-N+16}
                              {N(N+1)^2(N+2)^2}S_3\N\\
&&           -\frac{2}{3}\frac{N+4}
                              {(N+1)(N+2)^2}\zeta_3
             +\frac{10N^5+48N^4+122N^3+222N^2+213N+64}
                   {2N(N+1)^3(N+2)^3}S_2\N\\
&&           +\frac{2N^5+48N^4+174N^3+242N^2+161N+64}
                   {2N(N+1)^3(N+2)^3}S^2_1
             +\frac{4N^3+26N^2+51N+32}
                    {(N+1)^2(N+2)^3}\zeta_2\N\\
&&           -\frac{\PO_{26}S_1}
                    {N(N+1)^4(N+2)^4}
             +\frac{\PO_{27}}
                    {(N+1)^4(N+2)^5}
                            \Biggr\} ~,       \label{resPep} 
\end{eqnarray} \begin{eqnarray}
\PO_{26}&=&8N^7+120N^6+595N^5+1538N^4+2432N^3+2419N^2+1329N+256~, \N\\
\PO_{27}&=&22N^7+266N^6+1360N^5+3826N^4+6400N^3+6376N^2+3515N+832~. \N\\
\overline{A}^{Qg}_s&=&T_FC_A\Biggl\{
     -\frac{2}{3}\frac{3S_3+\zeta_3}{N^2(N+1)^2}
     +\frac{2N^3+N^2-3N-1}
           {N^3(N+1)^3}\Bigl(2S_2+\zeta_2\Bigr)\N\\
&&   +\frac{\PO_{28}}
           {N^5(N+1)^5(N+2)^2}
                            \Biggr\} ~,       \label{resSep} \\
\PO_{28}&=&4N^9+8N^8+6N^7+35N^6+66N^5-6N^4-85N^3-61N^2-24N-4~. \N\\
%
\overline{A}^{Qg}_t&=&T_FC_A\Biggl\{
    \frac{2}{3}\frac{N^2+3N+4}
          {(N-1)N(N+1)^2(N+2)}\Bigl(3S_3+\zeta_3\Bigr)\N\\
&& -\frac{2N^4+5N^3-3N^2-20N-16}
          {(N-1)N(N+1)^3(N+2)^2}\Bigl(2S_2+\zeta_2\Bigr)
   -\frac{\PO_{29}}
         {(N-1)N(N+1)^5(N+2)^4}
                            \Biggr\} ~,       \label{resTep} \\
\PO_{29}&=&2N^8+10N^7+28N^6+91N^5+213N^4+160N^3-248N^2-512N-256~. 
\\
%
A^{Qg}_{u}&=&
         a_s^2S^2_{\ep}
         \Biggl\{\frac{8}{3\ep}T_F
         \Bigl(1+\frac{\zeta_2}{8}\ep^2
\Bigr)\sum_{i=1}^3\Biggl(\frac{m_i^2}{\mu^2}\Biggr)^{\ep/2}\Biggr\}
         \Biggl\{
         -8T_F\Biggl(\frac{m^2}{\mu^2}\Biggr)^{\ep/2}
         \Biggl(
         \frac{1}{\ep}\frac{N^2+3N+2}{N(N+1)(N+2)} \N\\
       &&-\frac{1}{(N+1)(N+2)}
         +\ep\Bigl(\frac{1}{2(N+1)(N+2)}+\frac{\zeta_2}{8N}\Bigr)\Biggr)
          \Biggr\}~,\\ \N\\
%
A^{Qg}_{v}&=&
         a_s^2S^2_{\ep}
         \Biggl\{\frac{8}{3\ep}T_F
         \Bigl(1+\frac{\zeta_2}{8}\ep^2
\Bigr)\sum_{i=1}^3\Biggl(\frac{m_i^2}{\mu^2}\Biggr)^{\ep/2}\Biggr\}
         \Biggl\{
        16T_F\Biggl(\frac{m^2}{\mu^2}\Biggr)^{\ep/2}  \N\\
      && \times\Biggl(
\frac{1}{\ep}-\frac{1}{2}+\ep\left(\frac{1}{4}+\frac{\zeta_2}{8}\right)
        \Biggr)\frac{1}{(N+1)(N+2)}
        \Biggr\}~.
\end{eqnarray}
Note that for diagram $u$ and $v$ the sum runs over all heavy quark 
flavors.

The diagrams contributing to the pure singlet contributions yield
\begin{eqnarray}
%
%
\overline{A}^{Qq}_{a}&=&
        T_FC_F\Biggl\{
                - \frac{4}{3}\frac{(N+2)(N-1)}
                  {N^2(N+1)^2}\Bigl(3S_3+\zeta_3\Bigr)
                -2\frac{5N^3-5N^2-16N-4}
                  {N^3(N+1)^3(N+2)}\Bigl(2S_2+\zeta_2\Bigr)\N\\
&&              + \frac{2\PO_{30}}
                       {N^5(N+1)^5(N+2)^3}
               \Biggr\}~, \label{resAQqep} \\
\PO_{30}&=& 5N^9+25N^8-35N^7-229N^6-107N^5+481N^4
             +688N^3+360N^2+112N+16~.   \N\\
%
\overline{A}^{Qq}_{b}&=&
        T_FC_F\Biggl\{
           -\frac{32}{3}\frac{3S_3+\zeta_3}
                             {(N-1)N(N+1)(N+2)}
           -32\frac{2N+3}
                   {(N-1)N(N+1)^2(N+2)^2}\Bigl(2S_2+\zeta_2\Bigr)\N\\
&&         +32\frac{(2N+3)(N^4+6N^3+5N^2-12N-16)}
                   {N(N-1)(N+1)^4(N+2)^4}
               \Biggr\}~. \label{resBQqep} 
\end{eqnarray}

The flavor non--singlet contributions are given by
\begin{eqnarray}
%
%
\overline{A}_a^{qq,Q}&=&
              T_FC_F\Biggl\{
                 -\frac{2}{9}\frac{(N+2)(N-1)}
                                  {N(N+1)}\zeta_3
                 -\frac{2}{9}\frac{N^4+2N^3-10N^2-5N+3}
                                  {N^2(N+1)^2}\zeta_2\N\\
&&               -\frac{2}{81}\frac{\PO_{31}}
                                   {N^4(N+1)^4}
               \Biggr\}~,\label{resAqqQep} \\
\PO_{31}&=&49N^8+196N^7-83N^6-533N^5-374N^4-59N^3
         -33N^2+9N+27.   \N\\
%
\overline{A}_b^{qq,Q}&=&
              T_FC_F\Biggl\{
                  \frac{4}{3}S_4
                 +\frac{4}{3}S_2\zeta_2
                 -\frac{8}{9}S_1\zeta_3
                 -\frac{20}{9}S_3
                 -\frac{20}{9}S_1\zeta_2
                 +\frac{8}{9}\zeta_3
                 +\frac{112}{27}S_2
                 +\frac{8}{9}\zeta_2\N\\
&&               -\frac{656}{81}S_1
                 +\frac{392}{81}
                     \Biggr\}~, \label{resBqqQep} \\
  \overline{A}_c^{qq,Q}&=&
        T_FC_F\Biggl\{-\frac{\zeta_2}{2}-\frac{89}{72} \Biggr\}~.
                   \label{resCqqQep}
 \end{eqnarray}

\newpage
\section{\bf\boldmath 
Infinite Sums}

\vspace{1mm}\noindent
In this appendix we list a series of infinite sums which were needed in 
the present analysis and are newly calculated. In addition we made use of 
the sums in \cite{BBK1}. $\sigma_1$ is a symbol for $\sum_{k=1}^{\infty} 
(1/k)$ and the corresponding sums are divergent. The calculation was partly 
performed using integral representations, solving difference equations  and 
using the summation package {\tt Sigma}, see also
\cite{BBKS1}.
  \subsection{\bf\boldmath Weighted Beta functions}
   \begin{eqnarray}
    \sum_{i=1}^{\infty}
    \frac{B(N,i)}{(i+N+2)^3}
     &=&(-1)^N 
\Biggl[\frac{4S_{1,-2}(N+2)
                    +2S_{-3}(N+2)
                    +2\zeta_2S_1(N+2) 
                    +2\zeta_3
                    }
                    {N(N+1)(N+2)}\N\\
& & +\frac{
                    -6S_{-2}(N+2)
                    -3\zeta_2}
                    {N(N+1)(N+2)}\Biggr]
      +\frac{1}{N(N+1)(N+2)^2}~,
           ~
           \label{Beta2}\\
    \sum_{i=1}^{\infty}
    \frac{B(N,i)}{(i+N+3)^2}
     &=&6(-1)^N\frac{2S_{-2}(N+3)+\zeta_2}
                    {N(N+1)(N+2)(N+3)}
        +\frac{N^4+5N^3+13N^2+18N+13}
                    {N(N+1)^2(N+2)^2(N+3)^2}~,\N\\
      &&   ~
           \label{Beta3a}\\
    \sum_{i=1}^{\infty}
    \frac{B(N,i)}{(i+N+4)^2}
     &=&24(-1)^N\frac{2S_{-2}(N+4)+\zeta_2}
                    {N(N+1)(N+2)(N+3)(N+4)} \N\\
       &&+\frac{N^6+11N^5+54N^4+143N^3+213N^2+178N+100}
                    {N(N+1)^2(N+2)^2(N+3)^2(N+4)^2}~,
           ~
           \label{Beta3}\\
    \sum_{i=1}^{\infty}
    \frac{B(N,i)}{(i+1)^2}
     &=&\frac{-1+2N}{N}
        +(1-N)S_{1,2}(N)
        +\frac{-1+N+N^2}{N}S_2(N)\N\\
      &&+\Bigl({\frac{1-N-{N}^{2}}{N}}+(-1+N)S_1(N)\Bigr)\zeta_2
        +(1-N)\zeta_3
           ~
           \label{Beta39}\\
    \sum_{i=1}^{\infty}
    \frac{B(N,i)}{(i+4)}
     &=&
        -\frac{-144+300N-415N^2+241N^3-63N^4+6N^5}{144N^2}\N\\
&&      +\frac{(N-1)(N-2)(N-3)(N-4)}{24}\Bigl(\zeta_2-S_2(N)\Bigr)
           ~
           ~.\label{Beta41}
   \end{eqnarray}
  \subsection{\bf\boldmath Weighted Beta functions and harmonic sums}
   \begin{eqnarray}
    \sum_{i=1}^{\infty}
    \frac{B(N,i)}{i+1}S_1(i)
     &=&(N-1)S_3(N-1) 
        -(N-1)\zeta_3
        -S_2(N-1)
        +\zeta_2~,
        ~
           \label{Beta4}\\
    \sum_{i=1}^{\infty}
    \frac{B(N,i)}{i+2}S_1(i)
     &=&\Bigl(\zeta_3-S_3(N-1)\Bigr)\frac{(N-1)(N-2)}{2}
        +\Bigl(\zeta_2-S_2(N-1)\Bigr)\frac{N^2-3N+3}{2}\N\\
      &&+\frac{2-N}{2}~,
           \label{Beta5}\\
    \sum_{i=1}^{\infty}
    \frac{B(N,i)}{i+3}S_1(i)
     &=&\Bigl(-\zeta_3+S_3(N-1)\Bigr)\frac{(N-1)(N-2)(N-3)}{6} \N
\end{eqnarray} 
\begin{eqnarray}
      &&-\Bigl(\zeta_2-S_2(N-1)\Bigr)\frac{3N^3-18N^2+33N-22}{12}
        +\frac{(6N-13)(N-3)}{24}~,\N\\
      &&     
           \label{Beta6}\\
    \sum_{i=1}^{\infty}
    \frac{B(N,i)}{i+4}S_1(i)
     &=&
        -\frac{-864+2040N^2-3746N^3+2453N^4-675N^5+66N^6}
              {864N^3}\N\\
&&      +\frac{300-550N+385N^2-110N^3+11N^4}{144}\Bigl(\zeta_2-S_2(N)\Bigr)\N\\
&&      +\frac{(N-1)(N-2)(N-3)(N-4)}{24}\Bigl(\zeta_3-S_3(N)\Bigr)
           \label{Beta42}\\
    \sum_{i=1}^{\infty}
    \frac{B(N,i)}{i+N+3}S_1(i)
     &=&\frac{\zeta_2-S_2(N+2)}{N+3}
        +\frac{3N^6+15N^5+36N^4+51N^3+52N^2+36N+8}
              {N^3(N+1)^3(N+2)^3}~, \N\\
      &&     
           \label{Beta9}\\
    \sum_{i=1}^{\infty}
    \frac{B(N,i)}{i+N+4}S_1(i)
     &=&\frac{\zeta_2-S_2(N+3)}{N+4}
         +2\frac{2N^9+24N^8+129N^7+408N^6+854N^5+1270N^4}
              {N^3(N+1)^3(N+2)^3(N+3)^3} \N\\
      &&+2\frac{1405N^3+1158N^2+594N+108}
              {N^3(N+1)^3(N+2)^3(N+3)^3}~,
           \label{Beta10}\\
    \sum_{i=1}^{\infty}
    \frac{B(N,i)}{(i+N+1)^2}S_1(i)
     &=&(-1)^N\frac{2S_{1,-2}(N)+S_{-3}(N)+\zeta_2S_1(N)+\zeta_3}{N(N+1)}
        +\frac{\zeta_2-S_2(N)}{(N+1)^2}~, \N\\
      &&     
           \label{Beta8}\\
    \sum_{i=1}^{\infty}
    \frac{B(N,i)}{i+1}S_1(i+N)
     &=&(N-1)\Bigl(     
               2S_3(N) 
               +S_1(N)S_2(N)
               -\zeta_2S_1(N)
               -2\zeta_3 
             \Bigr) \N\\
      &&-\frac{N-1}{N}\Bigl(
                       S_2(N)
                       -\zeta_2
                      \Bigr)
        +\frac{1-N+N^2}
              {N^2}S_1(N)
        +\frac{1}{N^3}
           \label{Beta11}\\
    \sum_{i=1}^{\infty}
    \frac{B(N,i)}{i+2}S_1(i+N)
     &=&\frac{(N-1)(N-2)}{2}\Bigl(     
               -2S_3(N) 
               -S_1(N)S_2(N)
               +\zeta_2S_1(N)
               +2\zeta_3 
             \Bigr) \N\\
      &&+\frac{(N-2)(2N-1)}{2N}\Bigl(
                       S_2(N)
                       -\zeta_2
                      \Bigr)
        -\frac{-4+6N-7N^2+2N^3}
              {4N^2}S_1(N) \N\\
      &&+\frac{2-2N^2+N^3}
              {2N^3}
           ~.\label{Beta12}
   \end{eqnarray}
   \begin{eqnarray}
    \sum_{i=1}^{\infty}
    \frac{B(N,i)}{i+3}S_1(i+N)
     &=&\frac{(N-1)(N-2)(N-3)}{6}\Bigl(     
               2S_3(N) 
               +S_1(N)S_2(N)
               -\zeta_2S_1(N)
               -2\zeta_3 
             \Bigr) \N\\
      &&-\frac{(N-3)(3N^2-6N+2)}{6N}\Bigl(
                       S_2(N)
                       -\zeta_2
                      \Bigr)
\N\\ & &
        -\frac{-12+21N^2-19N^3+4N^4}{12N^3} \N\\
     && +\frac{36-66N+85N^2-39N^3+6N^4}
              {36N^2}S_1(N)
           \label{Beta13}\\
    \sum_{i=1}^{\infty}
    \frac{B(N,i)}{i+4}S_1(i+N)
     &=&
        \frac{144-340N^2+397N^3-150N^4+18N^5}{144N^3}\N\\
&&      -\frac{-144+300N-415N^2+241N^3-63N^4+6N^5}{144N^2}S_1(N)\N
\end{eqnarray} \begin{eqnarray}
&&      +\Bigl\{
              -\frac{(N-4)(2N-3)(1-3N+N^2)}{12N} \N\\
&&            +\frac{(N-1)(N-2)(N-3)(N-4)}{24}S_1(N)
         \Bigr\}\Bigl(\zeta_2-S_2(N)\Bigr)\N\\
&&      +\frac{(N-1)(N-2)(N-3)(N-4)}{12}\Bigl(\zeta_3-S_3(N)\Bigr)
           \label{Beta43}\\
    \sum_{i=1}^{\infty}
    \frac{B(N,i)}{(i+N+1)^2}S_1(i+N)
     &=&\frac{(-1)^{N+1}}{N(N+1)}\Bigl(
                                       4S_{1,-2}(N+1)
                                      -2S_{-3}(N+1)
\N\\    &&                                  -2S_1(N+1)S_{-2}(N+1)                                    +\zeta_2 S_1(N+1)  -\zeta_3
                                      -2S_{-2}(N+1)\nonumber\\ & &                                      -\zeta_2
                                \Bigr)
        +\frac{S_1(N+1)}{N(N+1)^2}
   +\frac{1}{N(N+1)^3}~, \N\\
&&
           \label{Beta14}\\
    \sum_{i=1}^{\infty}
    \frac{B(N,i)}{(i+N+2)^2}S_1(i+N)
     &=&\frac{(-1)^{N+1}}{N(N+1)(N+2)}\Bigl(
                                       8S_{1,-2}(N+2)
                                      -4S_{-3}(N+2) \N\\
&&                                    -4S_1(N+2)S_{-2}(N+2)
                                      +2\zeta_2S_1(N+2)
                                      -2\zeta_3
\N\\ & &
-10S_{-2}(N+2)-5\zeta_2
                                \Bigr) 
   +\frac{1+N+N^2}{N(N+1)^2(N+2)^2}S_1(N+2)\N\\ & &
        -\frac{1+7N+6N^2+N^3}{N(N+1)^3(N+2)^3}~,\\
    \sum_{i=1}^{\infty}
    \frac{B(N,i)}{(i+N+3)}S_1(i+N)
     &=&
        \frac{16N^3+12+40N+30N^2+6N^4+N^5}{N^3(N+1)^2(N+2)^2(N+3)}\N\\
     &&+\frac{85N^2+36+66N+69N^3+34N^4+9N^5+N^6}
              {N^2(N+1)^2(N+2)^2(N+3)^2}S_1(N)\N\\
     &&-6(-1)^N\Biggl(\frac{\zeta(2)+2S_{-2}(N)}{N(N+1)(N+2)(N+3)}\Biggr)
          ~,
           \label{Beta34}\\
    \sum_{i=1}^{\infty}
    \frac{B(N,i)}{(i+N+4)}S_1(i+N)
     &=&
        \frac{144+564N+564N^2+361N^3+180N^4+62N^5+12N^6+N^7}
             {N^3(N+1)^2(N+2)^2(N+3)^2(N+4)}\N\\
&&
  +\frac{424N^5+110N^6+16N^7+N^8}
              {N^2(N+1)^2(N+2)^2(N+3)^2(N+4)^2}S_1(N)\N\\
&&
  +\frac{576+1200N+1660N^2+1576N^3+1013N^4}
              {N^2(N+1)^2(N+2)^2(N+3)^2(N+4)^2}S_1(N)\N\\
     &&-24(-1)^N\Biggl(\frac{\zeta(2)+2S_{-2}(N)}{N(N+1)(N+2)(N+3)(N+4)}\Biggr)
          ~,
           \label{Beta35}
\\
    \sum_{i=1}^{\infty}
    B(N,i)S_1(i+N)^2
     &=&
{\frac {-2+5N-2{N}^{2}+{N}^{3}}{ (N-1) ^{3}{N}^{3}}}
+2{\frac { ( 1-N+{N}^{2} ) S_1(N)}{ (N-1) ^
{2}{N}^{2}}}+{\frac {  S_1(N)  ^{2}}{N-1}}\N\\
&&
+2{\frac {
 ( -1 ) ^{N}S_{-2}(N)}{ (N-1) N}}+{\frac {
 ( -1 ) ^{N}\zeta_2}{ (N-1) N}}
           \label{Beta30}
\\
    \sum_{i=1}^{\infty}
    \frac{B(N,i)}{i}S_1(i+N)^2
     &=&
\frac{2}{N^4}+2\frac{S_1(N)}{N^3}+\frac{S^2_1(N)}{N^2}
-S^2_1(N)S_2(N)-4S_1(N)S_3(N)\N\\
&&-3S_4(N)-2\frac{(-1)^NS_{-2}(N)}{N^2}
+2S^2_{-2}(N)-\frac{(-1)^{N}\zeta_2}{N^2}\N
\end{eqnarray}\begin{eqnarray}
&&+S^2_1(N)\zeta_2+2S_{-2}(N)
\zeta_2+\frac{17}{10}\zeta_2^2+4S_1(N)\zeta_3
           ~,
           \label{Beta32}\\
    \sum_{i=1}^{\infty}
    \frac{B(N,i)}{i+N}S_1(i+N)^2
     &=&
         \frac{(-1)^{N+1}}{N^2}\Bigl(\zeta_2+2S_{-2}(N)\Bigr)
        +\frac{S_1(N)^2}{N^2}
        +2\frac{S_1(N)}{N^3}
        +\frac{2}{N^4} 
           \label{Beta17} \\
    \sum_{i=1}^{\infty}
    \frac{B(N,i)}{i+N+1}S_1(i+N)^2
     &=&
         \frac{(-1)^{N}}{N(N+1)}\Bigl(
                                      -6S_{-2,1}(N-1)
                                      +2S_{-2}(N-1)S_1(N-1) \N\\
&&                                    +S_{-3}(N-1)
                                      +\zeta_2S_1(N-1)
                                      -3\zeta_3
                                      -2S_{-2}(N-1)
                                      -\zeta_2
                                \Bigr) \N\\
&&      +\frac{1+N+N^2}
              {N^2(N+1)^2}S_1(N-1)^2
        +2\frac{1+2N}
               {N^2(N+1)^2}S_1(N-1)
\N\\ & &        
+\frac{2+3N}
             {N^3(N+1)^2}~, \N\\
&&
           \label{Beta18}\\
     \sum_{i=1}^{\infty}
     \frac{B(N,i)}{i+N+2}S_1(i+N)^2
      &=&
          \frac{(-1)^{N}}{N(N+1)(N+2)}\Bigl(
                                       -12S_{-2,1}(N-1)
                                       +4S_{-2}(N-1)S_1(N-1) \N\\
&&                                    +2S_{-3}(N-1)
                                       +2\zeta_2S_1(N-1)
                                       -6\zeta_3
                                       +2(N-1)S_{-2}(N-1) \N\\
&&                                    +(N-5)\zeta_2
                                 \Bigr)
         +\frac{4+6N+7N^2+4N^3+N^4}
               {N^2(N+1)^2(N+2)^2}S_1(N-1)^2\N\\
&&      +4\frac{3+3N+N^2}
                {N^2(N+1)(N+2)^2}S_1(N-1)
         +\frac{4+11N+5N^2}
              {N^3(N+1)(N+2)^2}
         ~, \N\\
           \label{Beta19}\\
    \sum_{i=1}^{\infty}
    \frac{B(N,i)}{i+N}S_1(i)^2
     &=&\frac{S_{1,2}(N-1)-2S_3(N-1)-\zeta_2S_1(N-1)+3\zeta_3}{N}
        ~,           
           \label{Beta20}\\
    \sum_{i=1}^{\infty}
    \frac{B(N,i)}{i+N+1}S_1(i)^2
     &=&
        \frac{S_{1,2}(N)-2S_3(N)-\zeta_2S_1(N)+3\zeta_3}{N+1}
        -\frac{S_2(N)-\zeta_2}{N^2}+\frac{2}{N^4}
        ~,\N\\ &&    
          \label{Beta21}\\
    \sum_{i=1}^{\infty}
    \frac{B(N,i)}{i+N+2}S_1(i)^2
     &=&
        \frac{S_{1,2}(N+1)-2S_3(N+1)-\zeta_2S_1(N+1)+3\zeta_3}{N+2}\N\\
&&      -\frac{2N^2+N+1}{N^2(N+1)^2}\Bigl(S_2(N+1)-\zeta_2\Bigr)
\N\\ & &        
+\frac{5N^4+8N^3+13N^2+8N+2}{N^4(N+1)^4}~,    
           \label{Beta22}\\
    \sum_{i=1}^{\infty}
    \frac{B(N,i)}{i+N+2}S_1(i)S_1(N+i)
     &=&
        \frac{(-1)^N}{N(N+1)(N+2)}\Biggl(
                                        4S_{-2,1}(N)
                                       -6S_{-3}(N)
                                       -4S_{-2}(N)S_1(N)\N\\
&&                                     -2\zeta_2S_1(N)
                                       -2\zeta_3
                                       -2\frac{\zeta_2}{(N+1)}
                                       -4\frac{S_{-2}(N)}{(N+1)}
                                   \Biggr)
        -2\frac{S_3(N)}{N+2} \N\\
&& - \frac{S_1(N) S_2(N)}{N+2} + \frac{\zeta_2 S_1(N)}{N+2} +  
\frac{2 \zeta_3}{N+2} \N\\
&&      +\frac{2+7N+7N^2+5N^3+N^4}
              {N^3(N+1)^3(N+2)}S_1(N)
\N
   \end{eqnarray}
   \begin{eqnarray}
 &&      +2\frac{2+7N+9N^2+4N^3+N^4}
              {N^4(N+1)^3(N+2)}
        ~,\N\\ &&    
           \label{Beta25}
\\
    \sum_{i=1}^{\infty}
    B(N,i)S_2(N+i)
     &=&
{\frac {1}{ (N-1) {N}^{2}}}+{\frac {S_2(N)}{N-1}}-2{
\frac { ( -1 ) ^{N}S_{-2}(N)}{ (N-1) N}}-{
\frac { ( -1 ) ^{N}\zeta_2}{ (N-1) N}}\N\\
&&                   
           \label{Beta31}
\\
    \sum_{i=1}^{\infty}
    \frac{B(N,i)}{i}S_2(N+i)
     &=&
        -3S_4(N)
        -2S_{-2}(N)^2
        -S_2(N)^2
        -2S_{-2}(N)\zeta_2
        +S_2(N)\zeta_2
        +\frac{7}{10}\zeta_2^2  \N\\
&&      +(-1)^N \frac{\zeta(2)+2S_{-2}(N)}{N^2}
        +\frac{S_2(N)}{N^2}
        ~,           
           \label{Beta26}\\
    \sum_{i=1}^{\infty}
    \frac{B(N,i)}{N+i}S_2(N+i)
     &=&
        (-1)^N\frac{\zeta(2)+2S_{-2}(N)}{N^2}
        +\frac{S_2(N)}{N^2}
        ~,           
           \label{Beta27}\\
    \sum_{i=1}^{\infty}
    \frac{B(N,i)}{N+i+1}S_2(N+i)
     &=&
\frac{1+N+N^2}{N^2(N+1)^2}S_2(N)
+(-1)^N\Biggl[
-\frac{2S_{1,-2}(N)}{N(N+1)}
+\frac{2S_{-2}(N)}{N^2}\N\\
&&-\frac{S_{-3}(N)}{N(N+1)}
+\frac{\zeta_2}{N^2}
-\frac{S_1(N)\zeta_2}{N(N+1)}
-\frac{\zeta_3}{N(N+1)}
          \Biggr]
        ~,           
           \label{Beta33}\\
    \sum_{i=1}^{\infty}
    \frac{B(N,i)}{N+i+2}S_2(N+i)
     &=&
\frac{N^4+4N^3+7N^2+6N+4}
      {N^2(N+1)^2(N+2)^2}S_2(N)
+\frac{2}{N(N+1)^2(N+2)}\N\\
&&+(-1)^N\Biggl[
-\frac{4S_{1,-2}(N)}{N(N+1)(N+2)}
-\frac{2(N-2)S_{-2}(N)}{N^2(N+2)}\N\\
&&-\frac{2S_{-3}(N)}{N(N+1)(N+2)}
+\zeta_2\Bigl(-\frac{2S_1(N)}{N(N+1)(N+2)}-\frac{N-2}{N^2(N+2)}\Bigr)\N\\
&&-\frac{2\zeta_3}{N(N+1)(N+2)}
\Biggr]
        ~,           
           \label{Beta40}\\
    \sum_{i=1}^{\infty}
    \frac{B(N,i)}{N+i+2}S_2(i)
     &=&
{\frac{S_{1,2}(N)}{N+2}}
-{\frac{S_1(N)\zeta_2}{N+2}}+{\frac{\zeta_3}{N+2}}
-{\frac{(2+3N+4{N}^{2}+{N}^{3})S_2(N)}{{N}^{2}(N+1)^{2}(N+2)}} \N\\
&&+{\frac{(2+3N+4{N}^{2}+{N}^{3})\zeta_2}{{N}^{2}(N+1)^{2}(N+2)}}
+{\frac{2}{N(N+1)^{4}(N+2)}}
        ~,           
           \label{Beta28}
\\
   \sum_{i=1}^{\infty}
    \frac{B(N,i)}{(N+i+2)^2}S_1(i)
     &=&
{\frac{-8+N+3{N}^{2}}{N(N+1)^{4}(N+2)
^{2}}}
-{\frac{S_2(N)}{(N+2)^{2}}}
+{\frac{\zeta_2}{(N+2)^{2}}} \N\\
&& +\Bigl(
{\frac{2\zeta_2}{N(1+N)^{2}(N+2)}}
+{\frac{2S_1(N)\zeta_2}{N(N+1)(N+2)}} \N\\
&&+{\frac{2\zeta_3}{N(N+1)(N+2)}}
+{\frac{4S_{-2}(N)}{N(N+1)^{2}(N+2)}}\N\\
&&+{\frac{2S_{-3}(N)}{N(N+1)(N+2)}}
+{\frac{4S_{1,-2}(N)}{N(N+1)(N+2)}}
\Bigr)(-1)^{N}
        ~,     
        \label{Beta29} 
\end{eqnarray}
  \subsection{\bf\boldmath Weighted Harmonic Sums}
   \begin{eqnarray}
    \sum_{i=1}^{\infty}\frac{S_1(i)S_1(i+N)}{i+N}
     &=&
\frac{\sigma_1^3-\zeta_3-S^3_1(N)-2S_3(N)}{3}
+\frac{S^2_1(N)+S_2(N)}{N} -S_1(N)S_2(N)
        ~,
        \label{Harm3} 
\\
    \sum_{i=1}^{\infty}\frac{S_1(i)S_1(i+N)}{i+N+1}
     &=&
\frac{\sigma_1^3-4\zeta_3}{3}+\frac{S_3(N)-S^3_1(N)}{3}-S_1(N)\zeta_2
        ~,
        \label{Harm4} \\
    \sum_{i=1}^{\infty}\frac{S_1(i)S_1(i+N)}{i+N+2}
     &=&
\frac{\sigma_1^3-4\zeta_3}{3}
+\frac{S_3(N)-S^3_1(N)}{3}
-\frac{S^2_1(N)}{N+1}
-S_1(N)\zeta_2
-\frac{\zeta_2}{N+1} \N\\
&& -\frac{(N+2)}{(N+1)^2}S_1(N) 
-\frac{1}{(N+1)^2}
        ~,
        \label{Harm5} 
\\
    \sum_{i=1}^{\infty}\frac{S_1(i)S_1(i+N)}{i+3}
     &=&
\frac{\sigma_1^3-\zeta_3}{3}
+S_{1,1,1}(N)
-\frac{3N^2-6N+2}{N(N-1)(N-2)}S_{1,1}(N)
+\frac{13N-19}{4(N-1)(N-2)} \N\\
&&-\frac{(N+1)(7N-6)}{4N(N-1)}S_1(N)
          ~,
        \label{Harm6} \\
    \sum_{i=1}^{\infty}\frac{S_1(i)S_1(i+N)}{i+4}
     &=&
        \frac{1}{3}\sigma_1^3
        -\frac{(N+1)(132-232N+85N^2)}{36(N-2)(N-1)N}S_1(N)
        -\frac{1}{3}\zeta_3
        +S_{1,1,1}(N)\N\\
&&      -2\frac{(-3+2N)(1-3N+{N}^{2})}{(N-3)(N-2)(N-1)N}S_{1,1}(N)
        +\frac{809-909N+232N^2}{36(N-3)(N-2)(N-1)}
        \label{Harm67} \nonumber\\
\\
    \sum_{i=1}^{\infty}\frac{S_1(i+N)S_1^2(i)}{i}
     &=&
\frac{\sigma_1^4}{4}
+\frac{43}{20}\zeta_2^2
+3S_1(N)\zeta_3
+\frac{S^2_1(N)-S_2(N)}{2}\zeta_2
-S_1(N)S_{2,1}(N) \N\\
&&+\frac{S^2_1(N)S_2(N)}{2}
+\frac{2}{3}S_1(N)S_3(N)
-\frac{S^2_2(N)}{4}
+\frac{S^4_1(N)}{12}
~,
         \label{Harm26}\\
    \sum_{i=1}^{\infty}\frac{S^2_1(i+N)S_1(i)}{i}
     &=&
\frac{\sigma_1^4}{4}
+\frac{43}{20}\zeta_2^2
+5S_1(N)\zeta_3
+\frac{3S^2_1(N)-S_2(N)}{2}\zeta_2
-2S_1(N)S_{2,1}(N) \N\\
&&+S^2_1(N)S_2(N)
+S_1(N)S_3(N)
-\frac{S^2_2(N)}{4}
+\frac{S^4_1(N)}{4}
~,
         \label{Harm27}\\
    \sum_{i=1}^{\infty}\frac{S_1(i)S_2(i+N)}{i}
     &=&
\frac{\sigma_1^2}{2}\zeta_2
-\frac{1}{5}\zeta_2^2
-S_1(N)\zeta_3
-\frac{S^2_1(N)-3S_2(N)}{2}\zeta_2
-2S_{3,1}(N)
+\frac{1}{2}S_4(N) \N\\
&&+\frac{S^2_1(N)S_2(N)}{2}
+S_1(N)S_3(N)
~,
         \label{Harm30}\\
    \sum_{i=1}^{\infty}\frac{S_1(i+N)S_2(i)}{i}
     &=&
\frac{\sigma_1^2}{2}\zeta_2
-\frac{\zeta_2^2}{5}
+S_1(N)\zeta_3
+\frac{S^2_1(N)-S_2(N)}{2}\zeta_2
+2S_{1,1,2}(N)\N\\
&&-2S_{1,3}(N)
+S_1(N)S_{2,1}(N)
-S^2_1(N)S_2(N)
+\frac{S_4(N)-S^2_2(N)}{2}
~,
         \label{Harm32}\\
     \sum_{i=1}^{\infty}\frac{S_{1,1}(i+N)S_1(i)}{i+N}
      &=&
\frac{\sigma_1^4}{8}
+\frac{\sigma^2_1}{4}\zeta_2
-\frac{9}{40}\zeta^2_2
-3\frac{S_4(N)}{4}
+\frac{S^3_1(N)}{2N}
-\frac{S^4_1(N)}{8}
-3\frac{S^2_1(N)S_2(N)}{4}\N\\
&&-3\frac{S^2_2(N)}{8}
+S_1(N)\Bigl(3\frac{S_2(N)}{2N}-S_3(N)\Bigr)
+\frac{S_3(N)}{N}
~, \N\\
        ~
         \label{Harm49}
\end{eqnarray}\begin{eqnarray}
    \sum_{i=1}^{\infty}\frac{S_{1,1}(i+N)S_1(i)}{i}
     &=&
\frac{\sigma^4_1}{8}
+\frac{\sigma^2_1\zeta_2}{4}
+{\frac{39}{40}}\zeta^2_2
+2S_1(N)\zeta_3
+\frac{S^2_1(N)+S_2(N)}{2}\zeta_2
-S_{3,1}(N)\N\\
&&+3\frac{S^2_1(N)S_2(N)}{4}
+\frac{S^4_1(N)}{8}
-\frac{S^2_2(N)}{8}
+S_1(N)\Bigl(-S_{2,1}(N)+S_3(N)\Bigr)\N\\
&&+\frac{S_4(N)}{4}
~.
       ~
         \label{Harm50}\\
    \sum_{i=1}^{\infty}\frac{S_{1}(i)S_2(i+N)}{i+N}
     &=&
\frac{\sigma^2_1}{2}\zeta_2
-\frac{9}{10}\zeta^2_2
+\Bigl(-\frac{S_1(N)}{N}+\frac{S^2_1(N)}{2}+\frac{S_2(N)}{2}\Bigr)\zeta_2
-S^2_1(N)S_2(N)\N\\
&&-\frac{S^2_2(N)}{2}
-\frac{S_{2,1}(N)}{N}
+S_1(N)\Bigl(2{\frac{S_2(N)}{N}}+S_{2,1}(N)-2S_3(N)\Bigr)\N\\
&&+2\frac{S_3(N)}{N}
+S_{3,1}(N)
-\frac{3S_4(N)}{2}
~,
         \label{Harm55}\\
    \sum_{i=1}^{\infty}\frac{S_{2}(i)S_1(i+N)}{i+N}
     &=&
\frac{\sigma^2_1}{2}\zeta_2
-\frac{7}{10}\zeta^2_2
+\Bigl(\frac{2}{N}-2S_1(N)\Bigr)\zeta_3
+\Bigl(\frac{S_1(N)}{N}-\frac{S^2_1(N)}{2}-\frac{S_2(N)}{2}\Bigr)\zeta_2\N
\end{eqnarray}\begin{eqnarray}
&&-\frac{S^2_1(N)}{N^2}
-\frac{S_2(N)}{N^2}
+\frac{S^2_2(N)}{2}
-\frac{S_{2,1}(N)}{N}
+S_1(N)S_{2,1}(N)
+S_{3,1}(N)\N\\
&&+\frac{S_4(N)}{2}
~,
         \label{Harm56}\\
    \sum_{i=1}^{\infty}\frac{S_{1}(i+N)S_1(i)}{(i+N)^2}
     &=&
\frac{6}{5}\zeta^2_2
+\Bigl(-\frac{2}{N}+2S_1(N)\Bigr)\zeta_3
+\frac{S^2_1(N)}{N^2}
+\frac{S_2(N)}{N^2}
-\frac{S^2_2(N)}{2}
+\frac{S_{2,1}(N)}{N}\N\\
&&-S_1(N)S_{2,1}(N)
-S_{3,1}(N)
-\frac{S_4(N)}{2}
~,
         \label{Harm57}\\
    \sum_{i=1}^{\infty}\frac{S_{1}(i+N)S^2_1(i)}{i+N}
     &=&
\frac{\sigma^4_1}{4}
-\frac{3\zeta^2_2}{4}
+\Bigl(\frac{2}{N}-2S_1(N)\Bigr)\zeta_3
+\Bigl(\frac{S_1(N)}{N}-\frac{S^2_1(N)}{2}-\frac{S_2(N)}{2}\Bigr)\zeta_2\N\\
&&+\frac{S^3_1(N)}{N}
-\frac{S^4_1(N)}{4}
+S^2_1(N)\Bigl(-\frac{1}{N^2}-\frac{3S_2(N)}{2}\Bigr)
-\frac{S_2(N)}{N^2}
-\frac{S^2_2(N)}{4}\N\\
&&-\frac{S_{2,1}(N)}{N}
+S_1(N)\Bigl(3\frac{S_2(N)}{N}+S_{2,1}(N)-2S_3(N)\Bigr)
+2\frac{S_3(N)}{N}
+S_{3,1}(N)\N\\
&&-S_4(N)
~,
         \label{Harm58}
\end{eqnarray}
  \subsection{\bf\boldmath Harmonic Sums}
   \begin{eqnarray}
    \sum_{i=1}^{\infty}S_3(i+N)-S_3(i)
     &=&
       S_2(N)-(N+1)S_3(N)+N\zeta_3
        ~,
        ~
        \label{Harm34} 
\\
    \sum_{i=1}^{\infty}(S_3(i+N)-S_3(i))i
     &=&
\frac{S_1(N)-S_2(N)+N(N+1)S_3(N)-N(N+1)\zeta_3}{2}\N\\
&&-NS_2(N)+N\zeta_2 ~,
        ~
        \label{Harm35} \\
    \sum_{i=1}^{\infty}\Bigl(S_1(i+N)-S_1(i)\Bigr)^3
     &=&
-\frac{3}{2}S^2_1(N)-S^3_1(N)-\frac{1}{2}
S_2(N)+3NS_{2,1}(N)-NS_3(N)\N\\
&&+N\zeta_3
        ~,
        ~
        \label{Harm37} 
\end{eqnarray} \begin{eqnarray}
    \sum_{i=1}^{\infty}\Bigl(S_1(i+N)-S_1(i)\Bigr)^3i
     &=&
\frac{N}{2}\Bigl(
-(N+1)\zeta_3
-3(N+1)S_{2,1}(N)
+(N+1)S_3(N) \N\\
&&+(3N-1)\zeta_2
\Bigr)
+\frac{1+2N+6N^2}{4}S_2(N)
+\frac{3}{4}S^2_1(N)\N\\
&&+\frac{2-6N}{4}S_1(N)
        ~,
        ~
        \label{Harm46} \\
    \sum_{i=1}^{\infty}\Bigl(S_1(i+N)-S_1(i)\Bigr)S_2(i)
     &=&
-\frac{1}{2}S^2_1(N)-\frac{1}{2}S_2(N)+NS_{2,1}(N)+N\zeta_2-N
S_1(N)\zeta_2\N\\
&&+N\sigma_1\zeta_2-2N\zeta_3
        ~,
        ~
        \label{Harm38} \\
    \sum_{i=1}^{\infty}\Bigl(S_1(i+N)-S_1(i)\Bigr)S_2(i+N)
     &=&
 -(1+S_1(N))S_2(N)+NS_3(N)-NS_1(N)\zeta_2\N\\
&&+N(\zeta_2+\sigma_1\zeta_2-\zeta_3 ) 
        ~,
        ~
        \label{Harm39} 
   \end{eqnarray}
  \subsection{\bf\boldmath Miscellaneous Sums}
   \begin{eqnarray}
    \sum_{k=0}^{l-1}\frac{B(k+2,\ep/2)}{k+1}
     &=&\frac{4}{\ep^2}-\frac{2}{\ep}B(1+l,\ep/2)
        ~,\\
    \sum_{l=0}^{N-1}\binom{N-1}{l}(-1)^lB(l+1-\ep/2,2-\ep/2)
     &=&B(1-\ep/2,N+1-\ep/2)
        ~.
   \end{eqnarray}
  \subsection{\bf\boldmath Double and Other Sums}
  \begin{eqnarray}
   \sum_{i,j=1}^{\infty}\frac{S_1(i)S_1(i+j+N)}{i(i+j)(j+N)}
    &=&
6\frac{S_1(N)}{N}\zeta_3
+\zeta_2\Bigl(
        2\frac{S^2_1(N)}{N}+\frac{S_2(N)}{N}
        \Bigr)
+\frac{S^4_1(N)}{6N}
+\frac{S^2_1(N)S_2(N)}{N}\N\\
&&-\frac{S^2_2(N)}{N}
+4\frac{S_{2,1,1}(N)}{N}
+S_1(N)\Bigl(
       -3\frac{S_{2,1}(N)}{N}+4\frac{S_3(N)}{3N}
       \Bigr)\N\\
&&-2\frac{S_{3,1}(N)}{N}
-\frac{S_4(N)}{2N}~,
       \label{DoubleSum1}\\
   \sum_{k=1}^{\infty}\frac{B(k+\ep/2,N)}{N+k}       &=&\frac{1}{N^2}\N\\
       & +&\frac{\ep}{2}\Biggl\{ (-1)^N\frac{2S_{-2}(N)+\zeta_2}{N}
                       -\frac{S_1(N)}{N^2}
                        \Biggr\}\N\\
       & +&\ep^2\Biggl\{
                       -(-1)^N\frac{S_{-2,1}(N)}{N}
                       +(-1)^N\frac{2S_{-3}(N)-\zeta_3}{4N}\N\\
&&               +(-1)^N\frac{2S_{-2}(N)+\zeta_2}{4N}S_1(N)
                       +\frac{S^2_1(N)+S_2(N)}{8N^2}
                \Biggr\}\N
\end{eqnarray}\begin{eqnarray}
       & +&\frac{\ep^3}{48N}\Biggl\{
                        \frac{12}{5}(-1)^{N}\zeta^2_2
                        -6(-1)^NS_1(N)\zeta_3
                        +\zeta_2\Bigl(3(-1)^NS^2_1(N)\N\\
&&                      +3(-1)^NS_2(N)\Bigr)
                        +48(-1)^NS_{-2,1,1}(N)
                        -24(-1)^NS_{-3,1}(N)\N\\
&&                      -24(-1)^NS_{-2,2}(N)
                        +12(-1)^NS_{-4}(N)
                        +6(-1)^NS^2_1(N)S_{-2}(N)\N\\
&&                      +6(-1)^NS_2(N)S_{-2}(N)
                        -2\frac{S_3(N)}{N}
                        -\frac{S^3_1(N)}{N}\N\\
&&                      +S_1(N)\Bigl(
                                -3\frac{S_2(N)}{N}
                                -24(-1)^NS_{-2,1}(N)
                                +12(-1)^NS_{-3}(N)
                               \Bigr)
                              \Biggr\}\N\\
       &&+O(\ep^4)
~. 
\label{specialsum2}
\\
 \sum_{k=1}^{\infty}\frac{B(N+1,k+\ep/2)}{N+k}&=&
 (-1)^N\sum_{j=1}^{N}(-1)^j\Biggl(
  \sum_{k=1}^{\infty}\frac{B(j,k+\ep/2)}{j+k}
  +\frac{B(j,1+\ep/2)}{j}
                     \Biggr)
\N\\ & &
  +(-1)^N\sum_{k=1}^{\infty}\frac{B(1,k+\ep/2)}{k}~,
\end{eqnarray}
\begin{eqnarray}
   \sum_{k=1}^{\infty}\frac{B(k+\ep/2,N+1)}{N+k}
       &=&
          (-1)^N\Bigl[2S_{-2}(N)+\zeta_2\Bigr] \N\\
       &+&\frac{\ep}{2}(-1)^N\Bigl[
          -\zeta_3+\zeta_2S_1(N)+2S_{1,-2}(N)-2S_{-2,1}(N)
          \Bigr] \N\\
       &+&\frac{\ep^2}{4}(-1)^N\Biggl[
           \frac{2}{5}\zeta_2^2-\zeta_3S_1(N)+\zeta_2S_{1,1}(N)\N\\
       && +2\Bigl\{S_{1,1,-2}(N)+S_{-2,1,1}(N)-S_{1,-2,1}(N)\Bigr\}
          \Biggr]\N\\
       &+& \ep^3(-1)^N\Biggl[
                     -\frac{\zeta_5}{8}+\frac{S_1(N)}{20}\zeta_2^2
                     -\frac{S_{1,1}(N)}{8}\zeta_3
                     +\frac{S_{1,1,1}(N)}{8}\zeta_2\N\\
                   &&+\frac{S_{1,-2,1,1}(N)+S_{1,1,1,-2}(N)-S_{-2,1,1,1}(N)
                            -S_{1,1,-2,1}(N)}{4}
                    \Biggr]\N\\
       &&+O(\ep^4)~. \label{specialsum1}
       ~~
\end{eqnarray}
  \subsection{\bf\boldmath Expansion of harmonic sums for small argument}
One may expand nested harmonic sums into Taylor series w.r.t. the outer 
argument, using the corresponding differentiation rules \cite{STRUCT}. 
In the present calculation we made use of the following relations.
   \begin{eqnarray}
    S_1(\ep)&=&\zeta_2\ep-\zeta_3\ep^2+\frac{2}{5}\zeta_2^2\ep^3-\zeta_5\ep^4
              +O(\ep^5)~,
\\
    S_2(\ep)&=&2\zeta_3\ep-\frac{6}{5}\zeta^2_2\ep^2+4\zeta_5\ep^3
              -\frac{8}{7}\zeta_2^3\ep^4+O(\ep^5)~, 
\\
    S_3(\ep)&=&\frac{6}{5}\zeta^2_2\ep-6\zeta_5\ep^2+\frac{16}{7}\zeta^3_2\ep^3
              -15\zeta_7\ep^4+O(\ep^5)~,
\end{eqnarray}\begin{eqnarray}
    S_4(\ep)&=&4\zeta_5\ep-\frac{16}{7}\zeta^3_2\ep^2+20\zeta_7\ep^3
              -\frac{24}{5}\zeta^4_2\ep^4+O(\ep^5)~,\\ 
    S_{2,1}(\ep)&=&\frac{7}{10}\zeta^2_2\ep
                   +\Bigl(2\zeta_3\zeta_2-\frac{11}{2}\zeta_5\Bigr)\ep^2
                   +O(\ep^3) ~,\\ 
    S_{3,1}(\ep)&=&\Bigl(\frac{9}{2}\zeta_5-\zeta_3\zeta_2\Bigr)\ep
                   +O(\ep^2)~,\\ 
    S_{2,1,1}(\ep)&=&\Bigl(\frac{11}{2}\zeta_5-2\zeta_3\zeta_2\Bigr)\ep
                   +O(\ep^2)~. 
   \end{eqnarray}
These relations can be obtained expanding the representation of 
the sums in terms of Mellin transforms of Nielsen integrals 
weighted by $1/(1 \pm  x)$. In case of the single harmonic sums 
the expansions result from Euler's $\psi$--function and its 
derivatives. 

\subsection{\bf\boldmath Sample Calculation for one of the Sum}
\label{sec:samp}

\vspace{2mm}\noindent
In the following we illustrate the calculation of  sum~\eqref{DoubleSum1} in 
using the {\tt Sigma} package~:
\begin{eqnarray}
\sum_{j=1}^{\infty}\frac{1}{j+N}\sum_{i=1}^{\infty}\frac{S_1(i) S_1(i+j+N)}
{i (i+j)}.
\end{eqnarray}
First, we treat the inner sum for $N$ fixed,
\begin{eqnarray}
F(j)=\sum_{i=1}^{\infty}\frac{S_1(i) S_1(i+j+N)}{i (i+j)}~.
\end{eqnarray}
By {\tt Sigma}'s creative telescoping algorithm we compute the recurrence 
relation
\begin{eqnarray}
-(j+N+1)j^2F(j)+(j+1) \left(3 j^2+3 N j+7 j+2 N+4\right) F(j+1) 
& & 
\nonumber\\
-(j+2)\left(3 j^2+3 N j+11 j+4 N+10\right) F(j+2)
 & & 
\nonumber\\
  +(j+2) (j+3) (j+N+3) F(j+3)&=&A(j)+B(j)S_1(j+N)
\nonumber\\
\end{eqnarray}
where
\begin{eqnarray}
A(j)&=&\frac{1}{\left(j^2+3 j+2\right) \left(j^2+(2 N+3) j+N^2+3
   N+2\right)^2} \nonumber\\
   && \times \Big(3 j^5+(9 N+26) j^4+\left(10 N^2+63 N+86\right) j^3
   +\left(5 N^3+49N^2+156 N+137\right) j^2 \nonumber\\
   &&+\left(N^4+12 N^3+74 N^2+163 N+106\right) 
   j+5N^3+34 N^2+61 N+32\Big),\\
B(j)&=&\frac{3 j^3+(4 N+13) j^2+\left(-N^2+11 N+18\right)
   j-N^3-2 N^2+7 N+8}{(j+1) (j+2) (j+N+1) (j+N+2)}.
\end{eqnarray}
Next, we apply {\tt Sigma}'s recurrence solver and obtain three linearly independent solutions of the homogeneous version of the recurrence:
$$\frac{1}{j},\,\,\,\frac{S_1(j+N)}{j},\,\,\,\frac{-jS_1(j)+jS_1(j+N)+1}{j^2(N+1)}$$
and one solution of the recurrence itself:
\begin{eqnarray}
p(j)&=&\frac{S_1(j+N)^3}{6 j}-\frac{S_1(j+N)^2}{2 j^2}+\left(-\frac{S_2(N)}{2
   j}+\frac{S_2(j+N)}{2 j}+\frac{3}{N j+j}\right)
   S_1(j+N)-\frac{S_3(N)}{3 j}\nonumber\\ &&
   +\frac{S_1(N)^2}{2 N j+2 j}+\frac{3 N+2}{j^2
   (N+1)^2}+\frac{(j-2 (N+1)) S_1(N)}{2 j^2 (N+1)^2}+\frac{S_2(N)}{2
   j^2}-\frac{S_2(j+N)}{2 j^2}+\frac{S_3(j+N)}{3
   j}\nonumber\\ &&
   +\frac{\sum _{i=1}^j \frac{S_1(i+N)}{i^2}}{j}-\frac{\sum _{i=1}^j
   \frac{S_1(i+N)}{i}}{j^2}-\frac{\sum _{i=1}^j \frac{S_1(i)
   S_1(i+N)}{i}}{j}+\frac{\sum _{i=1}^j \frac{S_2(i+N)}{i}}{2 j}+\frac{\sum _{i=1}^j \frac{S_1(i+N)^2}{i}}{2
   j}\nonumber\\ &&
   +S_1(j) \left(\frac{-3 N-2}{j
   (N+1)^2}+\frac{S_1(N)}{N j+j}-\frac{S_2(N)}{2 j}+\frac{\sum _{i=1}^j
   \frac{S_1(i+N)}{i}}{j}\right)
-\frac{\sum _{i=1}^j\frac{S_1(i+N)}{(i+N)^2}}{j}.
\end{eqnarray}
The function $F(j)$ is given by
\begin{eqnarray}
F(j)=a_1 \frac{1}{j}+a_2 
\frac{S_1(j+N)}{j}+a_3\frac{-jS_1(j)+jS_1(j+N)+1}{j^2(N+1)}+p(j)
\end{eqnarray}
for some properly chosen constants $a_1$, $a_2$ and $a_3$ which are free of 
$j$. Looking at the initial values  for $j=1,2,3$ of $F(j)$ we 
can conclude that
\begin{eqnarray}
a_1&=&-\frac{1}{6} S_1(N)^3-\frac{S_1(N)^2}{2 N+2}+\left(-\frac{1}{2}
   S_2(N)-\frac{1}{2 (N+1)^2}\right) S_1(N)-S_{2,1}(N)+2 \zeta_3,\\
a_2&=&\frac{1}{2}
   \bigg(-S_1(N)^2-\frac{2 S_1(N)}{N+1}+S_2(N)+\frac{2 \left((N+1)^2
   \zeta_2-1\right)}{(N+1)^2}\bigg),\\
a_3&=&\frac{1}{2} (N+1)S_1(N)^2+S_1(N)-\frac{3 N+2}{N+1}.
\end{eqnarray}
One obtains
\begin{eqnarray}
F(j)&=&-\frac{S_1(N)^3}{6 j}+\frac{S_1(N)^2}{2 j^2}+\left(\frac{1}{2 j
   N^2}-\frac{S_2(N)}{2 j}\right) S_1(N)+\frac{S_1(j+N)^3}{6
   j}-\frac{S_1(j+N)^2}{2 j^2}
\nonumber\\ &&
   +S_1(j)^2 \left(-\frac{S_1(j+N)}{2
   j}-\frac{1}{2 j N}\right)+\frac{S_2(N)}{2 j^2}-\frac{S_2(j+N)}{2
   j^2}-\frac{S_3(N)}{3 j}+\frac{S_3(j+N)}{3
   j}
\nonumber\\ &&
-\frac{S_{2,1}N)}{j}+\frac{\sum _{i=1}^j \frac{S_1(i)^2}{i+N}}{2
   j}+\frac{\sum _{i=1}^j \frac{S_1(i+N)}{i^2}}{2 j}-\frac{(j+N) \sum
   _{i=1}^j \frac{S_1(i+N)}{i}}{j^2 N}-\frac{\sum _{i=1}^j
   \frac{S_1(i+N)}{(i+N)^2}}{j}
\nonumber\\ &&
   +S_1(j) \left(-\frac{S_1(N)^2}{2
   j}+\frac{S_1(j+N)}{j N}-\frac{S_2(N)}{2 j}+\frac{\sum _{i=1}^j
   \frac{S_1(i+N)}{i}}{j}+\frac{1}{2 j N^2}\right)
\nonumber\\ &&
   +\frac{\sum _{i=1}^j
   \frac{S_1(i+N)^2}{i}}{2 j}+\frac{\sum _{i=1}^j \frac{S_2(i+N)}{i}}{2
   j}+S_1(j+N) \left(\frac{S_2(j+N)}{2 j}-\frac{\frac{1}{N^2}-2 \zeta_2}{2
   j}\right)+\frac{2 \zeta_3}{j}.
\nonumber\\ 
\end{eqnarray}
Finally, we look at the indefinite nested sum
\begin{equation}\label{Equ:SecondSum}
S(N,a)=\sum_{j=1}^a\frac{F(j)}{j+N}
\end{equation}
with 
\begin{eqnarray}
\lim_{a\to\infty}S(N,a)=\sum_{j=1}^{a}\sum_{j=1}^{\infty}
\frac{S_1(i) S_1(i+j+N)}{i (i+j)(j+N)}.
\end{eqnarray}
At this point we emphasize that the sum expression~\eqref{Equ:SecondSum} 
with the derived sum representation of $F(j)$ fits into the input class 
of {\tt Sigma}. Hence, we can apply {\tt Sigma}'s machinery again and arrive 
for $S(a,N)$ at the following sum representation
\begin{eqnarray}
S(a,N)&=&-\frac{S_1(N)^4}{8 N}+\frac{S_1(a+N) S_1(N)^3}{6 N}
+\left(\frac{S_2(a)}{4N}-\frac{S_2(N)}{4 N}+\frac{\zeta_2}{2 N}\right) 
S_1(N)^2
\nonumber\\
&&+\left(S_1(a+N)
   \left(\frac{S_2(N)}{2 N}-\frac{1}{2
   N^3}\right)-\frac{S_{2,1}(N)}{N}+\frac{2 \zeta_3}{N}\right)
   S_1(N)-\frac{S_1(a+N)^4}{24 N}
\nonumber\\
&&+\frac{S_2(N)^2}{8
   N}
-\frac{S_2(a+N)^2}{8 N}+\frac{\left(\sum _{i=1}^a
   \frac{S_1(i+N)}{i}\right)^2}{2 N}+S_1(a)^3 \left(-\frac{S_1(a+N)}{3
   N}-\frac{1}{3 N^2}\right)
\nonumber\\
&&+S_1(a)^2 \left(-\frac{S_1(N)^2}{4
   N}+\frac{S_1(a+N)}{2 N^2}-\frac{S_2(N)}{4 N}+\frac{1}{4
   N^3}\right)+S_2(a) \left(\frac{S_2(N)}{4 N}+\frac{1}{12
   N^3}\right)
\nonumber\\
&&+\frac{S_3(N)}{3 N^2}-\frac{S_3(a+N)}{3
   N^2}+\frac{S_4(N)}{4 N}-\frac{S_4(a+N)}{4 N}-\frac{\sum _{i=1}^a
   \frac{S_1(i)}{(i+N)^3}}{3 N}-\frac{\sum _{i=1}^a
   \frac{S_1(i)^2}{(i+N)^2}}{2 N}
\nonumber\\
&&+\left(\frac{S_1(a)}{2
   N}-\frac{S_1(a+N)}{2 N}+\frac{1}{2 N^2}\right) \sum _{i=1}^a
   \frac{S_1(i)^2}{i+N}-\frac{\sum _{i=1}^a \frac{S_1(i)^3}{i+N}}{6
   N}-\frac{\sum _{i=1}^a \frac{S_1(i+N)}{i^3}}{6
   N}
\nonumber\\
&&+\left(\frac{S_1(a)}{2 N}-\frac{S_1(a+N)}{2 N}\right) \sum _{i=1}^a
   \frac{S_1(i+N)}{i^2}
\nonumber\\
&&
+\left(-\frac{S_1(a)}{N}+\frac{S_1(a+N)}{N
   }+\frac{1}{N^2}\right) \sum _{i=1}^a
   \frac{S_1(i+N)}{(i+N)^2}
\nonumber\\
&&+\frac{\sum _{i=1}^a
   \frac{S_1(i+N)}{(i+N)^3}}{N}-\frac{\sum _{i=1}^a \frac{S_1(i)
   S_1(i+N)}{i^2}}{2 N}+\frac{\sum _{i=1}^a \frac{S_1(i)
   S_1(i+N)}{(i+N)^2}}{N}+\frac{\sum _{i=1}^a \frac{S_1(i)^2
   S_1(i+N)}{i+N}}{N}
\nonumber\\
&&+\left(\frac{S_1(a)}{2 N}-\frac{S_1(a+N)}{2
   N}-\frac{1}{N^2}\right) \sum _{i=1}^a \frac{S_1(i+N)^2}{i}-\frac{\sum
   _{i=1}^a \frac{S_1(i+N)^2}{(i+N)^2}}{N}
\nonumber\\
&&+\frac{2 \sum _{i=1}^a
   \frac{S_1(i+N)^3}{i}}{3 N}+\frac{\sum _{i=1}^a \frac{S_1(i+N)
   S_2(i)}{i}}{2 N}+\left(\frac{S_1(a)}{2 N}-\frac{S_1(a+N)}{2 N}\right)
   \sum _{i=1}^a \frac{S_2(i+N)}{i}
\nonumber\\
&&+\frac{\sum _{i=1}^a
   \frac{S_1(i) S_1(i+N)^2}{i}}{2 N}-\frac{\sum _{i=1}^a \frac{S_1(i)
   S_2(i+N)}{i}}{2 N}+\frac{\sum _{i=1}^a \frac{S_1(i+N)
   S_2(i+N)}{i}}{N}
\nonumber\\
&&+\frac{S_2(a+N) \left(1-3 N^2 \zeta_2\right)}{6
   N^3}+\frac{S_2(N) \left(3 N^2 \zeta_2-1\right)}{6 N^3} 
+\frac{S_1(a+N)}{N^2}
\nonumber\\
&&+\left(\sum _{i=1}^a \frac{S_1(i+N)}{i}\right)
   \Bigg(\frac{S_1(a)^2}{2
   N}+\left(-\frac{S_1(a+N)}{N}-\frac{1}{N^2}\right)
   S_1(a)
   -\frac{S_1(N)^2}{2 N}
\nonumber
\\
&&-\frac{S_2(a)}{2
   N}-\frac{S_2(N)}{2 N}+\frac{\zeta_2}{N}-\frac{1}{2
   N^3}\Bigg)+S_1(a+N)^2
   \left(\frac{1-N^2 \zeta_2}{2 N^3}-\frac{S_2(a+N)}{4 N}\right)
\nonumber
\end{eqnarray}\begin{eqnarray}
&&+S_1(a+N)
   \left(\frac{S_3(N)}{3 N}-\frac{S_3(a+N)}{3
   N}+\frac{S_{2,1}(N)}{N}-\frac{2 \zeta_3}{N}\right)
\nonumber\\
&&+S_1(a)
   \Bigg(-\frac{S_1(N)^3}{6 N}+\frac{S_1(a+N) S_1(N)^2}{2
   N}+\left(\frac{1}{2 N^3}-\frac{S_2(N)}{2 N}\right) S_1(N)
\nonumber\\
&&+S_1(a+N)\left(\frac{S_2(N)}{2 N}-\frac{1}{2 N^3}\right)
-\frac{S_3(N)}{3
   N}+\frac{S_3(a+N)}{3 N}-\frac{S_{2,1}(N)}{N}+\frac{2 \zeta_3}{N}\Bigg).
\end{eqnarray}
We remark that all the sums in this expression are algebraically independent, i.e., no relations occur that could cancel some of the involved sums.

Finally, we send $a$ to infinity in the last expression 
and note that the involved sum expressions  can be simplified by the sum 
identities~\eqref{Harm3}--\eqref{Harm58} and some additional identities of 
similar type. In the final expression divergences of the type $\sigma_1^k$
being contained in some of the terms vanish. We find the right hand side 
of~\eqref{DoubleSum1}.

\newpage
\section{\bf\boldmath 
The first moment of the operator matrix 
element}

\vspace{1mm}\noindent
After the analytic continuation from the {\sf even} values of $N$ to 
$N~\epsilon~{\bf C}$ is performed one may consider the limit $N \rightarrow 
1$. In this procedure the term $(1 + (-1)^N)/2$ equals to 1. At $O(a_s^2)$ the 
terms $\propto T_F C_A$ contain $1/z$ contributions in momentum fraction space
and their first moment diverges. For the other contributions to the 
un--renormalized operator matrix element (after mass renormalization 
to 2--loop order), the first moment is related to the Abelian part of the 
transverse contribution to the gluon propagator $\Pi_V(p^2,m^2)|_{p^2=0}$, 
Figure~3,
\begin{figure}[h]
\begin{center}
\includegraphics[angle=0, width=4.0cm]{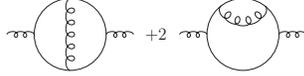} 
\end{center}
\caption{\label{fig:3}
\sf Abelian part  of the gluon self-energy due to heavy quarks. }
\end{figure}
except the term $\propto T_F^2$ which results from wave function 
renormalization. This  was shown in 
\cite{BUZA} up to the 
constant term in $\varepsilon$. One obtains
\begin{eqnarray}
\Pi_V(p^2,m^2) = S_\ep a_s T_F \Pi_V^{(1)}(p^2,m^2) 
+ S_{\ep}^2 
a_s^2 C_F T_F 
\Pi_V^{(2)}(p^2,m^2) + O(a_s^3)~, 
\end{eqnarray}
with
\begin{eqnarray}
\label{eqPI1}
\lim_{p^2 \rightarrow 0} \Pi_V^{(1)}(p^2,m^2) &=& \frac{1}{2} 
\hat{\hat{A}}_{Qg}^{(1), N=1}\\
\label{eqPI2}
\lim_{p^2 \rightarrow 0} \Pi_V^{(2)}(p^2,m^2) &=& \frac{1}{2} 
\hat{\hat{A}}_{Qg}^{(2), N=1}|_{C_F}~.
\end{eqnarray}
Here we extend the relation to the linear terms in $\ep$.  For the first 
moment
the double pole contributions in $\ep$ vanish in (\ref{eqPI1},\ref{eqPI2}).
From the corresponding QED-expressions $\Pi_T^{V (k)}$ given in \cite{DJOU} by 
asymptotic expansion of the photon propagator $(1/p^2) \tilde{\Pi}_V^{(k)}(p^2,m^2)$ 
in $m^2/p^2$ and adjusting the relative color factor 
for $k=2$ to $1/4 = 1/(C_F C_A)$, due to the transition from QED to QCD, the 
comparison can be performed up to the constant term
in $\ep$. One obtains
\begin{eqnarray}
\label{eqrPI1}
\lim_{p^2 \rightarrow 0} \frac{1}{p^2}
\tilde{\Pi}_V^{(1)}(p^2,m^2) &=& \frac{1}{2 T_F} \hat{\hat{A}}_{Qg}^{(1), N=1}
= - \left(\frac{m^2}{\mu^2}\right)^{\ep/2} S_\ep 
\left[\frac{8}{3 \ep} + \frac{\ep}{3} \zeta_2 \right]
\\
\label{eqrPI2}
\lim_{p^2 \rightarrow 0} \frac{1}{p^2}
\tilde{\Pi}_V^{(2)}(p^2,m^2) &=& \frac{1}{2 T_F C_F} \hat{\hat{A}}_{Qg}^{(2), 
N=1}|_{C_F}
= \left(\frac{m^2}{\mu^2}\right)^{\ep} 
\left[ - \frac{4}{\ep} +15 - \left(\frac{31}{4} + \zeta_2\right) \ep \right]~.
\end{eqnarray}
The latter term is easily obtained using {\tt MATAD} \cite{MATAD}.

\vspace{5mm}\noindent
{\bf Acknowledgments.}~~We would like to thank  K.~Chetyrkin, F.~Jegerlehner and 
M.~Steinhauser for useful conversations. We thank P. Paule for his interest in this work.
This work was supported in part by DFG Sonderforschungsbereich Transregio 9, 
Computergest\"utzte Theoretische Teilchenphysik, Spezialforschungsbereich F1305, 
project P20347-N18 of the Austrian FWF, and Studienstiftung des Deutschen Volkes. 
We used the code {\tt axodraw} \cite{axo} to draw Feynman diagrams. 

\newpage

\end{document}